\documentstyle[twocolumn,seceq,epsbox]{jpsj}

\def\runtitle{Theory of Electric Transport in the Pseudogap State 
of High-$T_{{\rm c}}$ Cuprates} 
\def\runauthor{Youichi {\sc Yanase}}

\title{ 
Theory of Electric Transport in the Pseudogap State 
of High-$T_{{\rm c}}$ Cuprates
} 

\author{Youichi {\sc Yanase}\footnote{E-mail: yanase@hosi.phys.s.u-tokyo.ac.jp}}

\inst{Department of Physics, University of Tokyo, Tokyo 113-0033}

\recdate{August 7, 2001}

\abst
{
 We investigate the electric transport in the pseudogap state 
of High-$T_{{\rm c}}$ cuprates. 
 Starting from the repulsive Hubbard model, we perform the microscopic 
calculation to describe the pseudogap phenomena which are induced by 
the superconducting fluctuations. 
 The single particle Green function, spin susceptibility 
and superconducting fluctuations are self-consistently determined 
by the SC-FLEX+T-matrix approximation. 
 The longitudinal and transverse conductivities are calculated 
by using the $\acute{{\rm E}}$liashberg and Kohno-Yamada formalism. 
 The effects of the spin fluctuations and superconducting fluctuations 
are estimated, respectively. 
 The vertex corrections arising from the two fluctuations are also calculated. 
 The complicated relations between the various effects are 
discussed in details. 
 The additional contribution from the Aslamazov-Larkin term is also 
estimated beyond the $\acute{{\rm E}}$liashberg formalism. 
 It is shown that the main effect of the superconducting fluctuations 
is the feedback effect through the spin fluctuations. 
 Therefore, the correct results are obtained by considering the 
superconducting fluctuations and the spin fluctuations simultaneously. 
 As a result, the temperature and doping dependences of 
the resistivity and the Hall coefficient are explained. 
 We point out that the characteristic momentum dependence of the systems 
plays an essential role in this explanation. 
}

\kword
{High-$T_{{\rm c}}$ cuprates; pseudogap; superconducting fluctuation; 
spin fluctuation; vertex correction; \\ resistivity; Hall coefficient; 
}

\begin{document}
\sloppy
\maketitle

\newcommand{\eli}{$\acute{{\rm E}}$liashberg }
\newcommand{\k}{\mbox{\boldmath$k$}}
\newcommand{\q}{\mbox{\boldmath$q$}}
\newcommand{\Q}{\mbox{\boldmath$Q$}}
\newcommand{\kk}{\mbox{\boldmath$k'$}}
\newcommand{\e}{\varepsilon}
\newcommand{\ee}{\varepsilon^{'}}
\newcommand{\s}{{\mit{\it \Sigma}}}
\newcommand{\J}{\mbox{\boldmath$J$}}
\newcommand{\vv}{\mbox{\boldmath$v$}}

\section{Introduction}

  The main issue of this work is the pseudogap phenomena in 
High-$T_{{\rm c}}$ cuprates which have attracted interests for many years.

  First, the pseudogap was found in the magnetic excitation channel 
by the nuclear magnetic resonance (NMR) experiment.~\cite{rf:yasuoka} 
 At present, the pseudogap phenomena have been observed in various quantities
which include 
NMR,~\cite{rf:yasuoka,rf:NMR}
neutron scattering,~\cite{rf:neutron} 
transport,~\cite{rf:iyereview,rf:takagi,rf:kubo,rf:ito,rf:takenaka,rf:haris,
rf:odatransport,rf:satoM,rf:hwang,rf:xiong,rf:mizuhashi,rf:ong} 
optical spectrum,~\cite{rf:homes} 
electronic specific heat,~\cite{rf:momono}
density of states~\cite{rf:renner}  
and the single particle spectral weight.~\cite{rf:ARPES}  
 The experimental results are reviewed in ref. 20.

 In our theory, the strong superconducting (SC) fluctuations are the origin 
of the pseudogap. 
 In other words, our theory belongs to the pairing fluctuation mechanism 
which has a broader 
sense.~\cite{rf:emery,rf:randeriareview,rf:janko,rf:haussman,
rf:micnas,rf:dagotto,rf:geshkenbein,rf:yanasePG,rf:jujo,rf:yanaseMG,
rf:yanaseSC,rf:onoda,rf:dahmpseudogap,rf:metzner,rf:kobayasi,rf:koikegami,
rf:yanaseFLEXPG,rf:koikegamiNSR,rf:kobayashiNSR,rf:stintzing}   
 The conventional weak coupling theory has concluded that the fluctuations 
are usually negligible in the superconducting phase transition. 
 Moreover, the effects on the single particle properties are less divergent 
compared with those on the two-body correlation function,~\cite{rf:AL,rf:MT} 
and usually neglected.
 However, the strong fluctuations naturally appear and sufficiently 
affect on the single particle properties when the strong coupling 
superconductivity occurs in the quasi-two dimensional 
systems.~\cite{rf:yanasePG} 
 Here, the coupling of the superconductivity is expressed by a parameter 
$T_{{\rm c}}^{{\rm MF}}/\varepsilon_{{\rm F}}$, where  
 $T_{{\rm c}}^{{\rm MF}}$ is the critical temperature in the mean field theory 
and $\varepsilon_{{\rm F}}$ is the effective Fermi energy. 
 The strong coupling superconductivity appears for a relatively large 
value of $T_{{\rm c}}^{{\rm MF}}/\varepsilon_{{\rm F}}$. 
 Since the high critical temperature and the strong electron correlation 
which reduces $\varepsilon_{{\rm F}}$ are the characteristics of 
the High-$T_{{\rm c}}$ compounds, 
 it is natural to consider a strong coupling superconductivity 
in these systems. 
 Indeed, the experimental results show the remarkably short coherence length 
$\xi_{0}$ which is proportional to 
$(T_{{\rm c}}^{{\rm MF}}/\varepsilon_{{\rm F}})^{-1}$ in the clean limit. 
 The short coherence length and the quasi-two 
dimensionality are the sufficient conditions of our 
theory.~\cite{rf:yanasePG} 

 We wish to mention that our theory is different from that of 
Nozi$\grave{{\rm e}}$res and Schmitt-Rink (NSR) 
theory.~\cite{rf:Nozieres,rf:tokumitu} 
 The NSR theory is one of the strong coupling theory 
describing the crossover from the BCS superconductivity 
to the Bose Einstein condensation. There are many works describing the 
pseudogap state as the crossover 
region.~\cite{rf:randeriareview,rf:haussman,rf:janko,
rf:koikegamiNSR,rf:kobayashiNSR,rf:stintzing} 
 However, it has been asserted that the NSR theory is justified only 
in the low density limit and not in High-$T_{{\rm c}}$ cuprates 
which are high density systems.~\cite{rf:yanasePG} 
 Moreover, the theory based on the NSR theory becomes harder in the strongly 
correlated systems. In this case, the superconductivity arises from 
the coherent quasi-particles near the Fermi surface. 
 This situation will be incompatible
with the NSR theory in which the chemical potential shifts lower than 
the bottom of the band.

 The realistic scenario is the resonance scattering scenario~\cite{rf:janko}
in which the self-energy correction gives rise to the pseudogap in the 
single particle properties. 
 The much easier condition is needed for this scenario in the realistic 
situation.~\cite{rf:yanasePG} 
 In addition to the mechanism of the pseudogap, 
we have obtained the comprehensive understanding of 
the magnetic field effects~\cite{rf:yanaseMG} and the superconducting 
transition~\cite{rf:yanaseSC} by using the T-matrix and self-consistent 
T-matrix approximations. 

 Moreover, we have succeeded in deriving the pseudogap phenomena 
by starting from the repulsive Hubbard model.~\cite{rf:yanaseFLEXPG} 
 The pseudogap in the single particle excitation~\cite{rf:renner,rf:ARPES} 
and the magnetic properties measured by the NMR~\cite{rf:yasuoka,rf:NMR} 
and neutron scattering~\cite{rf:neutron} have been explained in details. 
 This microscopic theory has well reproduced the doping dependence of 
the pseudogap including the electron-doped case. 
 It is probably a strong evidence that the pseudogap phenomena are 
microscopically derived under the reasonable condition, such as the 
reasonable critical temperature $T_{{\rm c}} \sim 100 {\rm K}$ and so on. 
 The formalism is adopted in this paper, and explained in \S2.1.

 In this stage, one of the important and unsolved issues 
is the transport phenomena in the pseudogap region 
which we study in this paper. 
 The electric transport shows its peculiar properties in the pseudogap state. 
 Although the many of the measured quantities 
(especially the magnetic properties) are remarkably affected 
by the pseudogap, the resistivity shows only a slight downward deviation 
from the $T$-linear 
dependence.~\cite{rf:ito,rf:takenaka,rf:odatransport,rf:mizuhashi}
 On the other hand, the Hall coefficient makes more distinct response to the 
pseudogap. That is, the Hall coefficient shows a broad peak and decreases 
when the temperature approaches 
$T_{{\rm c}}$.~\cite{rf:ito,rf:satoM,rf:haris,rf:mizuhashi,rf:xiong,rf:ong}  
 The origin of the characteristic responses has been a challenging problem 
for the theory of the pseudogap. 
 In this paper, we show that our theory naturally explains 
the response of the transport coefficients to the pseudogap.

 So far, the transport phenomena above the pseudogap temperature 
have been explained from the nearly anti-ferromagnetic Fermi liquid 
theory.~\cite{rf:hlubina,rf:stojkovic,rf:yanaseTR,rf:kontani,rf:kanki} 
 The $T$-linear resistivity and the enhancement of the Hall coefficient 
have been explained. 
 The essence of the above theory is explained in \S3. 
 However, the transport phenomena in the pseudogap region 
have not been studied extensively. 
 In this paper, we adopt the description based on the nearly 
anti-ferromagnetic Fermi liquid, and investigate the effects of the 
SC fluctuations.

 The electric transport has been a central issue of the 
theories of the SC fluctuations.~\cite{rf:AL,rf:MT,rf:dorin,rf:bieri,rf:IAV,rf:varlamov,rf:dorsey,rf:ikeda,rf:aronov}   
 However, these theories mostly discuss the $s$-wave case 
within the weak coupling theory, and therefore the calculations can be 
applied only in the narrow region near the critical point (or under the 
magnetic field~\cite{rf:ikeda,rf:dorsey,rf:aronov}). 
 The pseudogap phenomena induced by the 
SC fluctuations have been left out of view. 
 Therefore, these theories are not satisfactory for describing the 
pseudogap state. 
 Although some authors have investigated the effects 
through the single particle self-energy,~\cite{rf:ioffe} 
the qualitatively inconsistent results are obtained, as is explained later. 
 Thus, it is an important issue to investigate the effects of the 
SC fluctuations on the quasi-particle's transport, systematically. 
 In particular, we have to estimate the effects of the anti-ferromagnetic (AF) 
spin fluctuations and SC fluctuations simultaneously.

 In this paper, we apply the formalism used in Ref. 36 and calculate 
the transport coefficients in the pseudogap state. 
 Because both spin- and SC fluctuations play important roles, 
there exist many cooperative and competitive effects in a complicated way. 
 We explain these effects respectively, and clarify the main effect of 
the SC fluctuations. 

 First, the calculation within the lowest order with respect to 
$1/\tau_{{\rm c}}$ is carried out, where $\tau_{{\rm c}}$ is the lifetime of 
quasi-particles around $\k = (\pi/2,\pi/2)$. 
 We estimate 
the longitudinal and transverse conductivities by using the 
\eli and Kohno-Yamada formalism which is explained in \S2.2. 
 Because the lifetime $\tau_{{\rm c}}$ is sufficiently large, 
the \eli and Kohno-Yamada formalism based on the Fermi liquid theory 
is justified in the pseudogap state. 
 Actually, it is shown that the transport phenomena in the pseudogap state are 
well explained within the above treatment, by considering 
the spin fluctuations and the SC fluctuations simultaneously.

 The other contribution carried by the fluctuating Cooper pairs (that is the 
Aslamazov-Larkin term) is also estimated and shown to be negligible 
in the main region of the pseudogap state. 
 The Aslamazov-Larkin term is higher order with respect to 
$1/\tau_{\rm c}$, however more singular with respect to the mass 
term of the superconductivity. Therefore, this contribution becomes important 
near $T_{{\rm c}}$. However, the region is narrow.

 Thus, it is shown that our theory is consistent with the experimental results 
and with the previous theories. 
 As a result, the comprehensive understanding of the transport phenomena 
is obtained in a consistent way.

\section{Theoretical Framework}

\subsection{Superconducting fluctuations and pseudogap phenomena}

 First, we explain the theoretical framework to describe the 
SC fluctuations and the pseudogap phenomena. 
 The formalism used in this paper is the same as that in Ref. 36. 
 We show a brief outline in this subsection. 
 Hereafter, we use the unit  $\hbar=c=k_{{\rm B}}=1$.

 We use the Hubbard model,  
\begin{eqnarray}
  \label{eq:model}
  H =  \sum_{\mbox{\boldmath$k$},s}  
        \varepsilon_{\mbox{{\scriptsize \boldmath$k$}}} 
c_{\mbox{{\scriptsize \boldmath$k$}},s}^{\dag} 
c_{\mbox{{\scriptsize \boldmath$k$}},s}  
 + U \sum_{\mbox{\boldmath$k$},\mbox{\boldmath$k'$},\mbox{\boldmath$q$}} 
c_{\mbox{{\scriptsize \boldmath$q$}}-\mbox{{\scriptsize \boldmath$k'$}},
\downarrow}^{\dag} 
c_{\mbox{{\scriptsize \boldmath$k'$}},\uparrow}^{\dag} 
c_{\mbox{{\scriptsize \boldmath$k$}},\uparrow} 
c_{\mbox{{\scriptsize \boldmath$q$}}-\mbox{{\scriptsize \boldmath$k$}},
\downarrow},  
\nonumber \\
\end{eqnarray}
 where the two-dimensional dispersion relation 
$\varepsilon_{\mbox{{\scriptsize \boldmath$k$}}}$ is given by the 
tight-binding model including the nearest- and 
next-nearest-neighbor hopping $t$, $t'$, respectively, 
\begin{eqnarray}
 \label{eq:dispersion}
    \varepsilon_{\mbox{{\scriptsize \boldmath$k$}}} = 
                  -2 t (\cos k_{x} +\cos k_{y}) + 
                  4 t' \cos k_{x} \cos k_{y} - \mu. 
\nonumber \\
\end{eqnarray}
 The parameter $2 t=1 $, $t'=0.25 t$ and the lattice constant $ a = 1$ are 
fixed. 
 These parameters reproduce the typical Fermi surface of High-$T_{{\rm c}}$ 
cuprates. 
 The hole-doping concentration is defined as $\delta=1-n$.

 For the calculation of the SC fluctuations, 
we have to derive the attractive interaction in the $d$-wave channel. 
 In this paper, we start from the FLEX approximation~\cite{rf:FLEX} 
in order to describe the electronic state and the pairing interaction 
arising from the many body effects. 
 This approximation is a conserving approximation,~\cite{rf:baym} 
and has been used to describe the systems with strong AF 
spin fluctuations.~\cite{rf:moriyaAD,rf:monthouxFLEX,rf:paoFLEX,
rf:dahmFLEX,rf:langerFLEX,rf:deiszFLEX,rf:koikegamiFLEX,rf:takimotoFLEX} 
 In this approximation, the pairing interaction is mainly mediated by the AF 
spin fluctuations.~\cite{rf:moriya,rf:monthoux}

 The momentum dependence of the quasi-particle's lifetime arising from the 
spin fluctuations is an important property for the transport phenomena, 
as is explained in the following sections. 
 The quasi-particles at the Hot spot (near $(\pi,0)$) are strongly 
scattered and those at the Cold spot (near $(\pi/2,\pi/2)$) are 
only weakly scattered. 
 This momentum dependence is well reproduced by the FLEX approximation. 
 The FLEX approximation also reproduces the $d$-wave 
superconductivity with an appropriate critical temperature 
($T_{{\rm c}} \sim 100{\rm K}$).~\cite{rf:monthouxFLEX,rf:paoFLEX,
rf:dahmFLEX,rf:langerFLEX,rf:deiszFLEX,rf:koikegamiFLEX,rf:takimotoFLEX}
 This is important because $T_{{\rm c}}$ is an important parameter determining 
the strength of the SC fluctuations.~\cite{rf:yanasePG,rf:yanaseFLEXPG}

 The self-energy in the FLEX approximation is expressed as 
\begin{eqnarray}
 {\mit{\it \Sigma}}_{{\rm F}} (\mbox{\boldmath$k$}, {\rm i} \omega_{n}) & = & 
 T \sum_{\mbox{\boldmath$q$},{\rm i} \Omega_{n}} 
 V_{\rm n} (\mbox{\boldmath$q$}, {\rm i} \Omega_{n})
 G (\mbox{\boldmath$k-q$}, {\rm i} \omega_{n} - {\rm i} \Omega_{n}). 
\nonumber \\
\end{eqnarray}
 Here, $ V_{\rm n} (\mbox{\boldmath$q$}, {\rm i} \Omega_{n}) $ is the normal 
vertex, 
\begin{eqnarray}
 V_{\rm n} (\mbox{\boldmath$q$}, {\rm i} \Omega_{n}) & = & 
 U^{2} [\frac{3}{2} \chi_{{\rm s}} (\mbox{\boldmath$q$}, {\rm i} \Omega_{n}) 
      + \frac{1}{2} \chi_{{\rm c}} (\mbox{\boldmath$q$}, {\rm i} \Omega_{n}) 
      - \chi_{0} (\mbox{\boldmath$q$}, {\rm i} \Omega_{n})], 
\nonumber \\
\end{eqnarray}
where $\chi_{{\rm s}} (\mbox{\boldmath$q$}, {\rm i} \Omega_{n})$ and 
$\chi_{{\rm c}} (\mbox{\boldmath$q$}, {\rm i} \Omega_{n})$ are the 
spin and charge susceptibility, respectively.  
\begin{eqnarray}
\chi_{{\rm s}} (\mbox{\boldmath$q$}, {\rm i} \Omega_{n}) & = & 
 \frac{\chi_{0} (\mbox{\boldmath$q$}, {\rm i} \Omega_{n})}
      {1 - U \chi_{0} (\mbox{\boldmath$q$}, {\rm i} \Omega_{n})}, 
\hspace{5mm}
\nonumber \\
\chi_{{\rm c}} (\mbox{\boldmath$q$}, {\rm i} \Omega_{n}) & = & 
 \frac{\chi_{0} (\mbox{\boldmath$q$}, {\rm i} \Omega_{n})}
      {1 + U \chi_{0} (\mbox{\boldmath$q$}, {\rm i} \Omega_{n})}.    
\end{eqnarray}
In the above expression, 
$\chi_{0} (\mbox{\boldmath$q$}, {\rm i} \Omega_{n})$ is the 
irreducible susceptibility, 
\begin{eqnarray}
  \chi_{0} (\mbox{\boldmath$q$}, {\rm i} \Omega_{n}) = 
  - T \sum_{\mbox{\boldmath$k$},{\rm i} \omega_{n}} 
  G (\mbox{\boldmath$k$}, {\rm i} \omega_{n}) 
  G (\mbox{\boldmath$k+q$}, {\rm i} \omega_{n} + {\rm i} \Omega_{n}), 
\nonumber \\
\end{eqnarray} 
%
%
%
%
%
%
 where $G (\mbox{\boldmath$k$}, {\rm i} \omega_{n}) $ is the dressed Green 
function $G (\mbox{\boldmath$k$}, {\rm i} \omega_{n}) = 
( {\rm i} \omega_{n} - \varepsilon_{\mbox{{\scriptsize \boldmath$k$}}} - 
{\mit{\it \Sigma}}_{{\rm F}} (\mbox{\boldmath$k$}, {\rm i} \omega_{n}))^{-1} $,
and they are determined self-consistently. 
 We self-consistently solve eqs. (2.3)-(2.6) by the numerical calculation.

 In the main part of the following calculation, 
we divide the first Brillouin zone into $128 \times 128$ lattice points 
for the numerical calculation, while we have used the $64 \times 64$ lattice 
points in the previous paper.~\cite{rf:yanaseFLEXPG} 
 We find that the $128 \times 128$ lattice points are necessary to suppress 
the finite size effect in calculating the transport coefficients. 
 On the contrary, $64 \times 64$ lattice points are sufficient for 
the electric state and the magnetic properties. 
 The main reason of the difference is that the electric transport is 
mainly determined by the quasi-particles at the Cold spot. 
 We also checked the accuracy by comparing 
the results with those of $256 \times 256$ points. 
 The error is smaller than $4 \%$ for the Hall coefficient, and 
is much smaller for the other quantities. It is confirmed that 
the temperature dependence is not affected by the finite size effect.

 The criterion for the superconducting long range order is given by 
the condition that the Dyson-Gor'kov equation has a non-trivial solution.  
 The critical temperature $T_{{\rm c}}$ is determined from  
the $\acute{{\rm E}}$liashberg equation which is the following 
eigenvalue equation, 
\begin{eqnarray}
  \lambda \phi (\mbox{\boldmath$k$}, {\rm i} \omega_{n}) & = & 
 - T \sum_{\mbox{\boldmath$p$},{\rm i} \omega_{m}} 
   V_{\rm a} (\mbox{\boldmath$k-p$}, {\rm i} \omega_{n} - {\rm i} \omega_{m}) 
\nonumber \\
   & \times &
   |G (\mbox{\boldmath$p$}, {\rm i} \omega_{m})|^{2}
   \phi (\mbox{\boldmath$p$}, {\rm i} \omega_{m}). 
\end{eqnarray}
 The maximum eigenvalue $\lambda_{{\rm max}}$ becomes the unity at the 
critical temperature. 
 The corresponding eigenfunction 
$\phi_{{\rm max}} (\mbox{\boldmath$p$}, {\rm i} \omega_{m})$ 
is the wave function of the Cooper pairs. 
 Here, the anomalous vertex 
$V_{\rm a} (\mbox{\boldmath$q$}, {\rm i} \Omega_{n})$ is expressed as 
\begin{eqnarray}
   V_{\rm a} (\mbox{\boldmath$q$}, {\rm i} \Omega_{n}) & = & 
 U^{2} [\frac{3}{2} \chi_{{\rm s}} (\mbox{\boldmath$q$}, {\rm i} \Omega_{n}) 
      - \frac{1}{2} \chi_{{\rm c}} (\mbox{\boldmath$q$}, {\rm i} \Omega_{n})] 
      + U. 
\nonumber \\
\end{eqnarray}
 In this paper, the symmetry of the superconductivity is always 
the $d_{x^{2}-y^{2}}$-wave. 
 The important properties of the 
nearly anti-ferromagnetic Fermi liquid~\cite{rf:moriya,rf:monthoux,
rf:moriyaAD,rf:stojkovic,rf:yanaseTR,rf:kanki,rf:kontani,rf:hlubina} 
are well reproduced in the FLEX approximation.

\begin{figure}[htbp]
  \begin{center}
   \epsfysize=3cm
    $$\epsffile{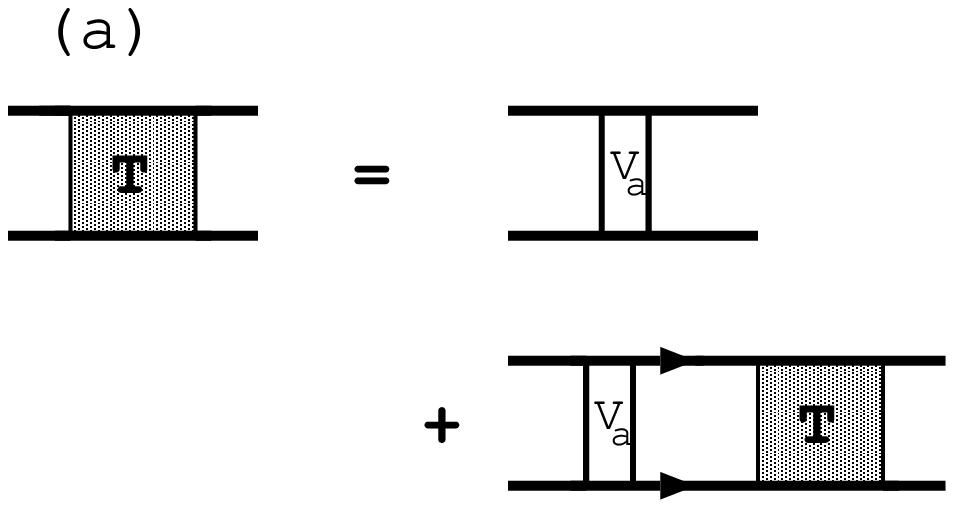}$$
\hspace{15mm}  
   \epsfysize=2.5cm
    $$\epsffile{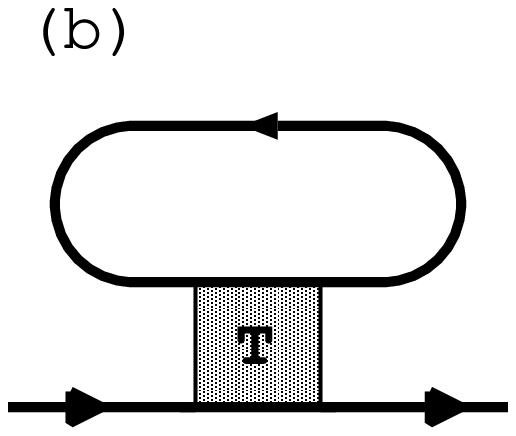}$$
    \caption{The diagrammatic descriptions of (a) the T-matrix and 
            (b) the self-energy arising from the SC fluctuations.}    
  \end{center}
\end{figure}

 In order to study the pseudogap phenomena induced by the SC fluctuations, 
we extend the FLEX approximation. 
 The SC fluctuations are generally represented by the 
T-matrix which is the propagator of the SC fluctuations. The T-matrix 
$T(\mbox{\boldmath$k_{1}$}, {\rm i} \omega_{n}: 
\mbox{\boldmath$k_{2}$}, {\rm i}  \omega_{m} : 
\mbox{\boldmath$q$}, {\rm i} \Omega_{n})$
is expressed by the ladder diagrams in the particle-particle channel 
(Fig. 1 (a)), and it is approximately estimated as,~\cite{rf:yanaseFLEXPG} 
\begin{eqnarray}
 &  &T(\mbox{\boldmath$k_{1}$}, {\rm i} \omega_{n}: 
\mbox{\boldmath$k_{2}$}, {\rm i}  \omega_{m} : 
\mbox{\boldmath$q$}, {\rm i} \Omega_{n}) 
\nonumber \\
 & &   =  \frac{ g \lambda(\mbox{\boldmath$q$}, {\rm i} \Omega_{n})
  \phi(\mbox{\boldmath$k_{1}$}, {\rm i} \omega_{n}) 
  \phi^{*}(\mbox{\boldmath$k_{2}$}, {\rm i} \omega_{m})} 
  {1 - \lambda(\mbox{\boldmath$q$}, {\rm i} \Omega_{n})}, 
\\
  \lambda(\mbox{\boldmath$q$}, {\rm i} \Omega_{n}) & = & 
  - T \sum_{\mbox{\boldmath$k$},{\rm i} \omega_{n}} 
      \sum_{\mbox{\boldmath$p$},{\rm i} \omega_{m}} 
  \phi^{*} (\mbox{\boldmath$k$}, {\rm i} \omega_{n})
   V_{\rm a} (\mbox{\boldmath$k-p$}, {\rm i} \omega_{n} - {\rm i} \omega_{m}) 
\nonumber \\
\times &G& (\mbox{\boldmath$p$}, {\rm i} \omega_{m}) 
   G (\mbox{\boldmath$q-p$}, {\rm i} \Omega_{n} - {\rm i} \omega_{m}) 
   \phi (\mbox{\boldmath$p$}, {\rm i} \omega_{m}), 
\end{eqnarray}
by extending the $\acute{{\rm E}}$liashberg equation. 

 Here, $\phi (\mbox{\boldmath$p$}, {\rm i} \omega_{m})$ is the eigenfunction 
of the $\acute{{\rm E}}$liashberg equation eq. (2.7) 
with its maximum eigenvalue $\lambda_{{\rm max}}$. 
 This function corresponds to the wave function 
of the fluctuating Cooper pairs. 
 The obtained momentum dependence of the wave function will be shown 
in Fig. 8. 
 The wave function is normalized as 
\begin{eqnarray}
  \sum_{\mbox{\boldmath$k$},{\rm i} \omega_{n}} 
|\phi(\mbox{\boldmath$k$}, {\rm i} \omega_{n})|^{2} = 1, 
\end{eqnarray}
and the constant factor $g$ is obtained as 
\begin{eqnarray}
     g = 
\sum_{\mbox{\boldmath$k_{1}$},{\rm i} \omega_{n}}
\sum_{\mbox{\boldmath$k_{2}$},{\rm i} \omega_{m}}
&\phi^{*}&(\mbox{\boldmath$k_{1}$}, {\rm i} \omega_{n})
V_{\rm a} (\mbox{\boldmath$k_{1}-k_{2}$}, 
{\rm i} \omega_{n} - {\rm i} \omega_{m}) 
\nonumber \\
&\times& \phi(\mbox{\boldmath$k_{2}$}, {\rm i} \omega_{m}). 
\end{eqnarray}

 In the above expression, the function $1-\lambda(\mbox{\boldmath$q$}, 0)$ 
shows the dispersion relation of the Cooper pairs with a finite momentum 
$\q$ of the center of mass. 
 We define the function $t(\mbox{\boldmath$q$}, {\rm i} \Omega_{n}) 
   =   g \lambda(\mbox{\boldmath$q$}, {\rm i} \Omega_{n})/
  (1 - \lambda(\mbox{\boldmath$q$}, {\rm i} \Omega_{n}))$ which 
corresponds to the T-matrix used in Ref. 32. 
 The mass term $t_{0}=1-\lambda(\mbox{\boldmath$0$}, 0)$ expresses 
the distance to the critical point. 
 The small $t_{0}$ means that the system is near the superconducting 
instability. 
 The above procedure is justified when the fluctuations are strong,  
that is, the parameter $t_{0}$ is small.

 The self-energy arising from the SC fluctuations is  
given as, 
\begin{eqnarray} 
 {\mit{\it \Sigma}}_{{\rm S}} (\mbox{\boldmath$k$}, {\rm i} \omega_{n}) =  
 &T& \sum_{\mbox{\boldmath$q$},{\rm i} \Omega_{n}} 
  T(\mbox{\boldmath$k$}, {\rm i} \omega_{n}: 
\mbox{\boldmath$k$}, {\rm i} \omega_{n}: 
\mbox{\boldmath$q$}, {\rm i} \Omega_{n}) 
\nonumber \\
&\times&
 G (\mbox{\boldmath$q-k$}, {\rm i} \Omega_{n} - {\rm i} \omega_{n}),  
\end{eqnarray} 
in the one-loop approximation (Fig. 1(b)). 
 The total self-energy is obtained by the summation, 
${\mit{\it \Sigma}} (\mbox{\boldmath$k$}, {\rm i} \omega_{n}) = 
{\mit{\it \Sigma}}_{{\rm F}} (\mbox{\boldmath$k$}, {\rm i} \omega_{n}) + 
{\mit{\it \Sigma}}_{{\rm S}} (\mbox{\boldmath$k$}, {\rm i} \omega_{n})$. 
 The pseudogap phenomena result from the self-energy correction 
${\mit{\it \Sigma}}_{{\rm S}} (\mbox{\boldmath$k$}, {\rm i} \omega_{n})$.~\cite{rf:yanaseFLEXPG} 
 Because of the momentum dependence of the wave function 
$\phi (\mbox{\boldmath$k$}, {\rm i} \omega_{n})$, the pseudogap has the similar
$d$-wave form to the superconducting gap. 
 In other words, the effects of the SC fluctuations are strong around  
$\k=(\pi,0)$ and weak around $\k=(\pi/2,\pi/2)$ on the Fermi 
surface. This is an important character of the pseudogap observed by 
ARPES.~\cite{rf:ARPES} 
 Therefore, the direct effects of the pseudogap on the transport phenomena 
are weak, since the electric transport is mainly carried 
by the quasi-particles at the Cold spot. 
 This momentum dependence plays an essential role in 
the transport phenomena in the pseudogap state.

 In the previous paper,~\cite{rf:yanaseFLEXPG} we have carried out both the 
lowest order calculation (FLEX+T-matrix approximation) and the self-consistent 
calculation (SC-FLEX+T-matrix approximation). 
 In the FLEX+T-matrix approximation, the functions obtained by the 
FLEX approximation are used in calculating eqs. (2.9)-(2.13). 
 In the SC-FLEX+T-matrix (SCFT) approximation, eqs. (2.3)-(2.13) are solved 
self-consistently.  
 The pseudogap phenomena are essentially obtained by 
the lowest order calculation. 
 However, we will show shortly that the feedback effect, which is not 
included in the FLEX+T-matrix approximation, is the main effect on  
the transport phenomena. 
 (This is the special character of the transport phenomena 
which are only indirectly affected by the pseudogap.) 
 Therefore, we perform the SC-FLEX+T-matrix approximation in this paper. 
 Although the effects of the SC fluctuations are weaker than those obtained 
by the FLEX+T-matrix approximation, the SCFT approximation gives the 
qualitatively same effects on the single particle and magnetic properties. 
 The characteristic results are shown in Fig. 2. 
 The density of state (DOS) shows the gap-like structure (Fig. 2(a)). 
 In addition to the results shown in the previous 
paper~\cite{rf:yanaseFLEXPG} ($U=2.4$), we show the results at $U=2.0$ which 
is the parameter used in this paper. 
 The pseudogap is more distinct in the under-doped and/or the strong 
interaction cases (Fig. 2(a) and the Fig. 19 in Ref. 36). 
 These properties are consistent with the experimental 
results.~\cite{rf:renner} 
 The magnetic properties are also affected by the SC 
fluctuations (Fig. 2(b)), although the peak of $1/T_{1}T$ is too close to the 
critical temperature ($T_{{\rm c}}=0.0033$ at $U=1.6$). 
 This is probably because the FLEX approximation overestimates the 
anti-ferromagnetic correlation and because the self-consistent calculation 
generally reduces the effects of the SC 
fluctuations.~\cite{rf:yanasePG} 
 However, the SC fluctuations reduces the $1/T_{1}T$ from much 
higher temperature than the mean field critical temperature 
$T_{{\rm c}}^{{\rm MF}}=0.0078$. 
 Since the position of the peak is determined 
by the competition between the spin and SC fluctuations, 
it is natural that the position is different between the different 
approximations. 
 Anyway, it is a general result of a theory of fluctuations 
that the effects of the fluctuations begin to appear above the mean field 
critical temperature $T > T_{{\rm c}}^{{\rm MF}}$.

\begin{figure}[htbp]
  \begin{center}
   \epsfysize=6.2cm
    $$\epsffile{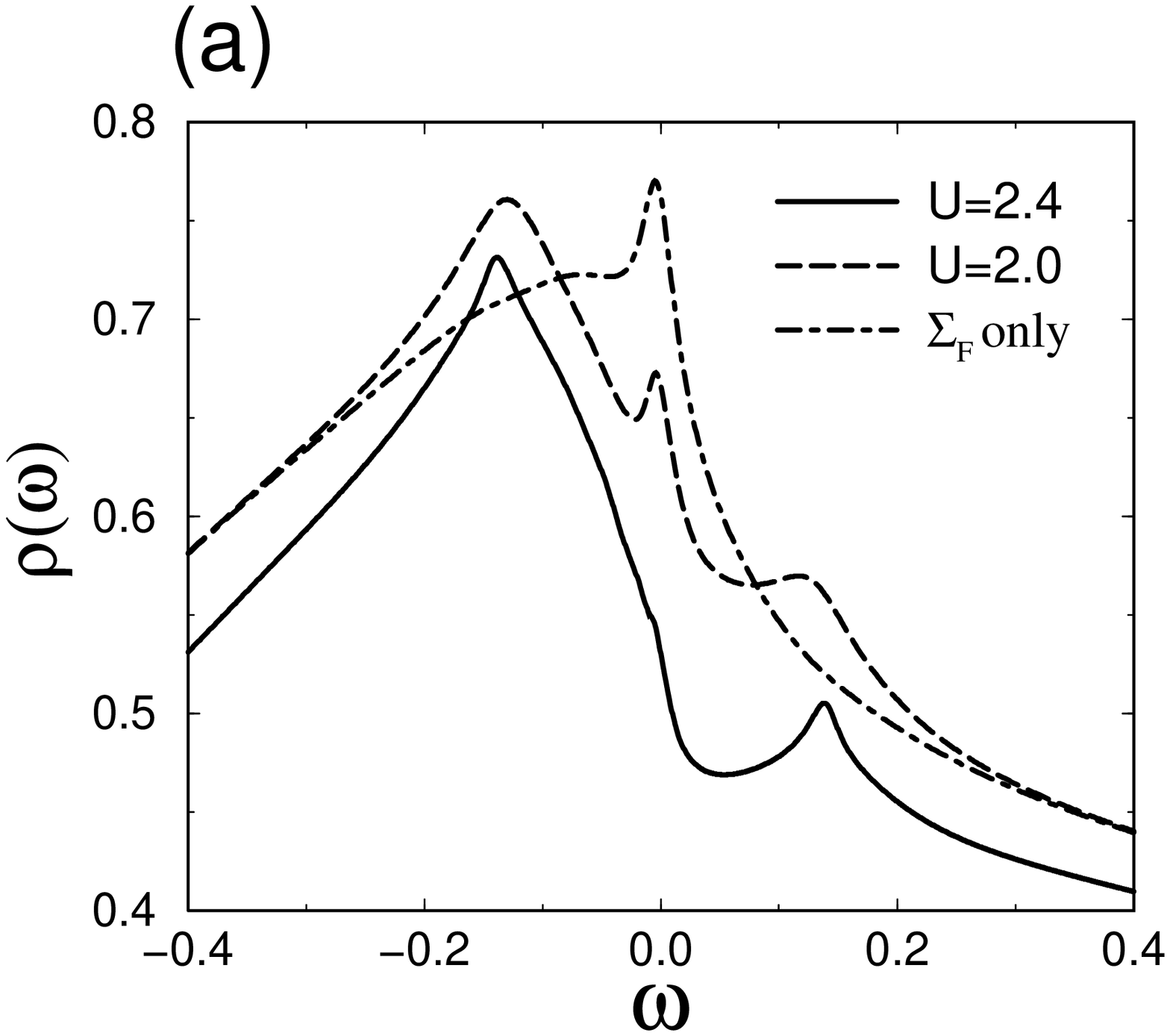}$$
   \epsfysize=5.8cm
    $$\epsffile{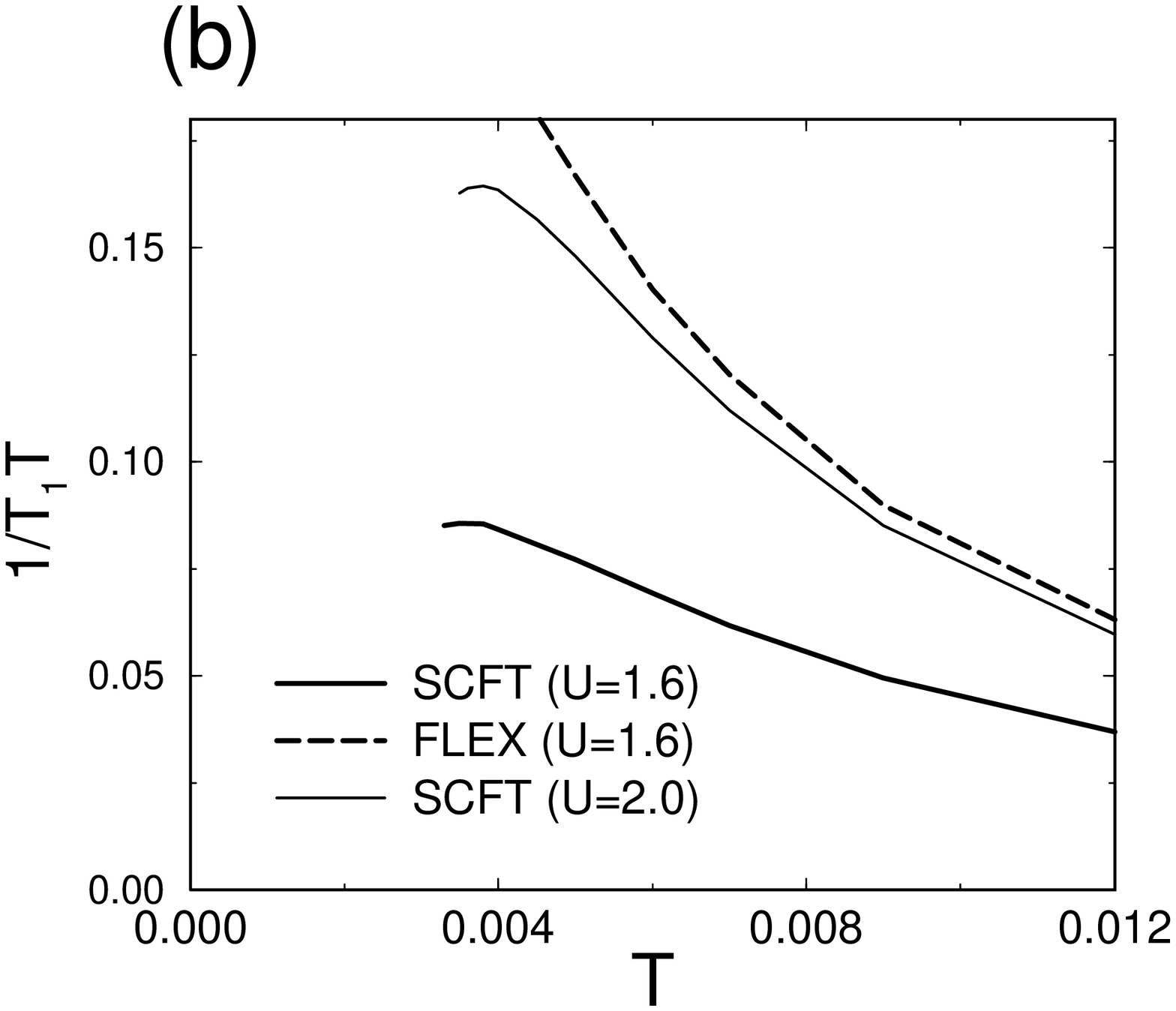}$$
    \caption{(a) The density of state obtained by the SCFT 
             approximation. The solid and long-dashed lines correspond to 
             the results $U=2.4$ and $U=2.0$, respectively. 
             The dash-dotted line show the results at $U=2.0$ 
             by neglecting the self-energy arising from the SC fluctuations 
             $\s_{{\rm S}}$. 
             (b) The NMR $1/T_{1}T$ obtained by the SCFT 
                approximation (thick solid line) and FLEX approximation 
                (long-dashed line) at $U=1.6$. 
                The results at $U=2.0$ 
                are shown by the thin solid line. 
             }    
  \end{center}
\end{figure}

\subsection{General theory for the electric transport}

 In this subsection, we review the general theory of the electric transport 
on the basis of the Fermi liquid theory. 
 In the Kubo formula, the electric conductivity is given by 
using the current-current correlation function. 
\begin{eqnarray}
  \sigma_{\mu\nu} & = & e^{2} {\rm lim}_{\omega=0} 
                    \frac{{\rm Im}K_{\mu\nu}^{{\rm R}}(\omega)}{\omega},
\\
  K_{\mu\nu}(\omega_{{\rm n}}) & = & \int_{0}^{\beta} {\rm d}\tau 
                       <T_{\tau} J_{\mu}(\tau) J_{\nu}(0)> 
                       e^{{\rm i} \omega_{{\rm n}} \tau}. 
\end{eqnarray}
 Here, $\omega_{{\rm n}} = 2 \pi n T $ is the bosonic 
Matsubara frequency, $T$ is the temperature ($\beta=1/T$), 
$e$ is the unit of charge ($e > 0$). 
 The current operator $J_{\mu}$ is defined as  
$J_{\mu} = \sum_{k, \sigma} v_{\mu}(\mbox{\boldmath$k$})
c_{\mbox{{\scriptsize \boldmath$k$}},\sigma}^{\dag} 
c_{\mbox{{\scriptsize \boldmath$k$}},\sigma} $ and 
$J_{\mu}(\tau) = e^{H \tau} J_{\mu} e^{-H \tau}$ where 
$v_{\mu}(\k) = \partial \e_{\k}/\partial k_{\mu} $ is the band velocity. 
 The correlation function $K_{\mu\nu}(\omega_{{\rm n}})$ can be rewritten as 
$K_{\mu\nu}(\omega_{{\rm n}}) = \sum_{\k,\sigma} \sum_{\kk,\sigma'}   
        v_{\mu}(\k) v_{\nu}(\kk) \int_{0}^{\beta} {\rm d}\tau 
       <T_{\tau} c_{\mbox{{\scriptsize \boldmath$k$}},\sigma}^{\dag} (\tau)
                 c_{\mbox{{\scriptsize \boldmath$k$}},\sigma} (\tau)
                 c_{\mbox{{\scriptsize \boldmath$k'$}},\sigma'}^{\dag}
                 c_{\mbox{{\scriptsize \boldmath$k'$}},\sigma'} > 
                       e^{{\rm i} \omega_{{\rm n}} \tau} $. 
 The retarded function $K^{{\rm R}}(\omega)$ is obtained from 
$K(\omega_{{\rm n}})$ by the analytical continuation 
${\rm i} \omega_{{\rm n}} \rightarrow \omega+{\rm i}\delta$.

 Generally speaking, the expression of $K^{{\rm R}}(\omega)$ is 
much complicated in the process of the analytic continuation. 
 However, \eli gave compact formula for the longitudinal conductivity 
$\sigma_{xx}$ 
by taking account of the most divergent terms with respect to the 
quasi-particle's lifetime 
$\tau(\mbox{\boldmath$k$}) = 1/\gamma(\mbox{\boldmath$k$})$ 
($\gamma(\mbox{\boldmath$k$}) = - {\rm Im} {\mit{\it \Sigma}}^{{\rm R}}
(\mbox{\boldmath$k$}, 0)$).  
 This procedure is based on the Fermi liquid theory and correct 
in the coherent limit 
$ v_{\mu}(\mbox{\boldmath$k$}) \tau(\mbox{\boldmath$k$}) |_{{\rm max}} \gg 1$
which is justified in the low temperature region. 
 The exceptional case is the system with a collective mode. 
 For example, the Aslamazov-Larkin (AL) term is higher order with respect 
to $1/v_{\mu}(\mbox{\boldmath$k$}) \tau(\mbox{\boldmath$k$})$, 
but divergent in the vicinity of the superconducting critical 
point.~\cite{rf:AL} 
 We calculate this term in \S5 and conclude that its contribution is not 
so important in our case. That is, the electric transport 
in the main part of the pseudogap region is explained 
within the \eli formalism. 

 The \eli formula is given by using the Green function, 
\begin{eqnarray}
 \sigma_{xx} = e^{2} \sum_{\k} \int \frac{{\rm d}\e}{\pi} (-f'(\e)) 
|G^{{\rm R}}(\k,\e)|^{2} \tilde{v}_{x}(\k,\e) J_{x}(\k,\e), 
\nonumber \\
\end{eqnarray}
where the function $f'(\e)$ is the first derivative of the Fermi distribution 
function and $\tilde{v}_{x}(\k,\e) = v_{x}(\k) + 
\partial {\rm Re} \s(\k,\e)/\partial k_{x}$ is the velocity including the 
$k$-mass renormalization. This renormalization corresponds to a part of 
the vertex correction. 
 The total current vertex $J_{x}(\k,\e)$ is obtained 
by solving the Bethe-Salpeter equation,  
\begin{eqnarray}
  J_{x}(\k,\e) &=& \tilde{v}_{x}(\k,\e) + \sum_{\kk} 
\int \frac{{\rm d}\ee}{4 \pi {\rm i}} \Im_{22}(\k,\e:\kk,\ee) 
\nonumber \\
&\times& |G^{{\rm R}}(\kk,\ee)|^{2} J_{x}(\kk,\ee). 
\end{eqnarray}
 The function $\Im_{22}(\k,\e:\kk,\ee)$ is obtained by the 
analytic continuation of the irreducible four point vertex function. 
 The explicit form is given in eq. (12) in Ref. 70. 
 The renormalization of the total current vertex $J_{x}(\k,\e)$ (eq. (2.17)) 
is generally the main contribution of the vertex correction. 
 Thus, in the following we use \lq vertex correction' as the renormalization 
arising from the vertex function $\Im_{22}(\k,\e:\kk,\ee)$.

 The vertex correction is necessary to satisfy the Ward identity which 
corresponds to the momentum conservation law.  
 For example, it can not be shown that the conductivity is infinite 
without Umklapp processes, unless the vertex correction is taken into 
account.~\cite{rf:yamada,rf:okabe,rf:maebashi} 
 Generally, the conductivity is finite in the lattice system 
with Umklapp processes. 
 Then, the effects of the vertex correction are usually taken into account 
by only multiplying a constant factor, and have no important role, 
qualitatively.~\cite{rf:yamada} 
 This argument is based on the assumption that the temperature dependence 
of the four point vertex is negligible, which is usually justified.  
 However, the vertex correction sometimes plays an important role  
when a collective mode induces a temperature dependence of the 
vertex.~\cite{rf:kontani,rf:kanki} 
 This is the case which we consider in this paper. 
 Indeed, we will show that the vertex correction is important for the 
transverse (Hall) conductivity $\sigma_{xy}$, while it is not so important 
for the longitudinal conductivity $\sigma_{xx}$.

 The expression for the Hall conductivity $\sigma_{xy}$ corresponding to 
eq. (2.16) was given by Kohno and Yamada.~\cite{rf:kohno} 
\begin{eqnarray}
  \sigma_{xy}  =  - H e^{3} \sum_{\k} \int \frac{{\rm d}\e}{\pi} (&-&f'(\e)) 
|{\rm Im} G^{{\rm R}}(\k,\e)| |G^{{\rm R}}(\k,\e)|^{2} 
\nonumber \\
\times \tilde{v}_{x}(\k,\e) 
[&J_{x}&(\k,\e) \partial J_{y}(\k,\e)/ \partial k_{y} 
\nonumber \\
&-& J_{y}(\k,\e) \partial J_{x}(\k,\e)/ \partial k_{y}].  
\end{eqnarray}
 In case of the Hall conductivity, the most divergent term with respect to 
$\tau(\mbox{\boldmath$k$})$ is the square term, 
$\sigma_{xy} \propto \tau(\mbox{\boldmath$k$})^{2}$. 
 This expression was obtained by calculating the current 
under the magnetic filed $H$ and estimating the linear term with respect to 
the field $H$.~\cite{rf:kohno} 
%
%
%
 In this paper, the magnetic field is fixed to be parallel to the {\it c}-axis 
$H \parallel {\it c}$, and the current $J$ is fixed to be perpendicular to the 
{\it c}-axis $J \perp {\it c}$.

 The above expressions (eqs. (2.16) and (2.18)) are rewritten to the more 
conventional form by using the formula, 
$ |G^{{\rm R}}(\k , \e )|^{2} = \pi \rho (\k , \e )/\gamma (\k , \e ) $ and 
$ |{\rm Im} G^{{\rm R}}(\k,\e)| |G^{{\rm R}}(\k,\e)|^{2} = 
\pi \rho (\k,\e)/2 \gamma (\k,\e)^{2}$. 
 Here, $\rho (\k,\e) = z(\k)\delta (\e-\varepsilon^{*}(\k))$ is 
the coherent part of the spectral weight, 
$z(\k) = 
(1 - \partial {\rm Re} \s^{{\rm R}} (\k,\e)/\partial \e|_{\e=\e^{*}})^{-1}$ 
is the mass renormalization factor and 
$\gamma(\k,\e) = -{\rm Im} \s^{{\rm R}} (\k,\e)$. 
 The energy of the quasi-particle $\varepsilon^{*}(\k)$  
is determined by the equation 
$\varepsilon^{*}(\k) - \varepsilon (\k) - {\rm Re} \s (\k,\e^{*}(\k)) = 0 $,
 which results in the conventional form 
$\varepsilon^{*}(\k) \cong z(\k) (\varepsilon (\k)+{\rm Re} \s (\k,0))$. 
 The resulting conductivities are expressed as, 
\begin{eqnarray}
  \sigma_{xx} & = & e^{2} \sum_{\k} z(\k) (-f'(\e^{*}(\k))) 
\tilde{v}_{x}(\k,\e^{*}(\k)) 
\nonumber \\
& & \times
J_{x}(\k,\e^{*}(\k))/\gamma(\k,\e^{*}(\k)),
\\
 & \cong & e^{2} \int_{{\rm FS}} \frac{{\rm d}k}{(2 \pi)^{2}} 
\frac{\tilde{v}_{x}(\k)}{\tilde{v}(\k)} J_{x}(\k) \tau(\k), 
\\
  \sigma_{xy} & = & -\frac{H e^{3}}{2} \sum_{\k} z(\k) (-f'(\e^{*}(\k))) 
\tilde{v}_{x}(\k,\e^{*}(\k)) 
\nonumber \\
& & \times 
[J_{x}(\k,\e^{*}(\k)) \frac{\partial J_{y}(\k,\e^{*}(\k))}{\partial k_{y}} 
\nonumber \\
- &J_{y}&(\k,\e^{*}(\k)) \frac{\partial J_{x}(\k,\e^{*}(\k))} {\partial k_{y}}]
/\gamma(\k,\e^{*}(\k))^{2},
\\
 & \cong &  \frac{H e^{3}}{4} \int_{{\rm FS}} \frac{{\rm d}k}{(2 \pi)^{2}} 
|J(\k)|^{2} (\frac{\partial \varphi(\k)}{\partial k_{\parallel}}) \tau(\k)^{2}.
\end{eqnarray}

 Here, the integration $\int_{{\rm FS}}$ is carried out on the Fermi 
surface. 
 Although the above expressions are similar to the results of Boltzmann 
equations, the velocity is replaced by the total current vertex $J_{\mu}$. 
 These expressions are justified in the Fermi liquid limit 
$z(\k) \gamma(\k) \ll T$. 
 In this limit, the electric conductivities are determined by the velocity 
$\tilde{v}(\k,0)$, the current vertex $J_{\mu}(\k,0)$ 
and the lifetime $\tau(\k)$. 
 Therefore, we have used the definition 
$\tilde{v}_{\mu}(\k)=\tilde{v}_{\mu}(\k,0)$, $J_{\mu}(\k)=J_{\mu}(\k,0)$
and so on.  
 In eq. (2.22), we use the angle of the current vertex which is defined as 
$\varphi(\k)={\rm Arctan}(J_{x}(\k)/J_{y}(\k))$ (See Fig. 5). 
 It is important that the Hall conductivity is proportional to the differential
of the angle $\varphi(\k)$ with respect to the momentum 
along the Fermi surface $k_{\parallel}$.

 In this paper, we use eqs. (2.16)-(2.18) in calculating the conductivities, 
although eqs. (2.19)-(2.22) are expected to give qualitatively same results. 
 This is because the Fermi liquid limit is not always justified and 
because the finite size effects are reduced by this procedure. 
 The resulting resistivity $\rho$ and Hall coefficient $R_{{\rm H}}$ 
are obtained by the formula, $\rho = 1/\sigma_{xx}$ and 
$R_{{\rm H}}=\sigma_{xy}/\sigma_{xx}^{2} H$, respectively. 
 Hereafter, we neglect the constant factor arising from the charge $e$.

 It should be noticed that the coherent transport is assumed in the above 
expressions. This assumption seems to be incompatible with the 
pseudogap induced by the large damping near the Fermi level. 
 However, this difficulty is removed by the characteristic momentum dependence 
of the systems, {\it i.e.}, the pseudogap occurs at the Hot spot, 
while the in-plane transport is determined by the Cold spot. 
 Because the coherency of the quasi-particles at the Cold spot 
is sufficiently maintained, the above formula are justified 
even in the pseudogap state.

 We comment on the {\it c}-axis transport which shows qualitatively different 
behaviors from the in-plane transport. 
 The {\it c}-axis resistivity shows a semi-conductive behavior in the 
pseudogap state.~\cite{rf:takenaka} 
 The {\it c}-axis optical conductivity $\sigma_{{\rm c}}(\omega)$ shows 
a pseudogap, while the in-plane optical conductivity 
$\sigma_{{\rm ab}}(\omega)$ shows a sharp Drude peak in the pseudogap 
state.~\cite{rf:homes} 
 Thus, the {\it c}-axis transport is incoherent while the in-plane transport 
is sufficiently coherent. 
 Since the formalism used in this paper assumes the coherent transport, 
the quantitative estimation for the c-axis transport is difficult. 
 However, we have obtained the consistent understanding also for the 
$c$-axis transport.~\cite{rf:yanaseTR,rf:ioffe}   
 The qualitative differences are explained from the characteristic momentum 
dependence of the matrix element of the inter-layer hopping 
$t_{\perp}(\k) \propto ({\rm cos} k_{x} - {\rm cos} k_{y})^{2}$ which was 
shown by the band calculation.~\cite{rf:okanderson} 
 In short, the {\it c}-axis transport is mainly determined by the Hot spot, 
and therefore the coherent transport is suppressed by the 
pseudogap.~\cite{rf:yanaseSC} 
 Thus, the incoherent {\it c}-axis transport in the pseudogap state 
is also explained in a consistent way.

\section{Electric Transport in the Nearly Anti-ferromagnetic Fermi Liquid}

 In this section, we review the effects of the AF 
spin fluctuations on the electric transport. 
 The detailed explanation has been given in the previous 
works.~\cite{rf:hlubina,rf:stojkovic,rf:yanaseTR,rf:kanki,rf:kontani} 
 Below, we explain the essential points and show the typical results 
obtained by the FLEX approximation. 
 The interaction is fixed to $U=1.6$, and the self-energy 
$\s(\k,\omega)=\s_{{\rm F}}(\k,\omega)$ 
is obtained by eqs. (2.3)-(2.6) in this section. 

 The corresponding four point vertex in the FLEX approximation 
is shown in Fig. 3(a-c).~\cite{rf:kontani} 
 Since the term (a) gives the dominant 
contribution,~\cite{rf:kontani} the terms (b) and (c) are neglected in this 
paper. 
 We call the term (a) spin fluctuation Maki-Thompson (SPMT) term 
in this paper. 
 The obtained vertex function in eq. (2.17) is expressed as, 
\begin{eqnarray}
  \Im_{22}^{{\rm F}}(\k,\e:\kk,\ee) &=& 
  2 {\rm i} ({\rm cth}(\frac{\e-\ee}{2 T}) + {\rm tanh}(\frac{\ee}{2 T})) 
\nonumber \\
 & &\times
  {\rm Im} V^{{\rm R}}_{{\rm n}} (\k-\kk,\e-\ee). 
\end{eqnarray}
 The current vertex $J(\k,\varepsilon)$ is calculated by solving eq. (2.17). 
The longitudinal and transverse conductivities are calculated
by using eqs. (2.16) and (2.18). 
\begin{figure}[htbp]
  \begin{center}
   \epsfysize=2.5cm
    $$\epsffile{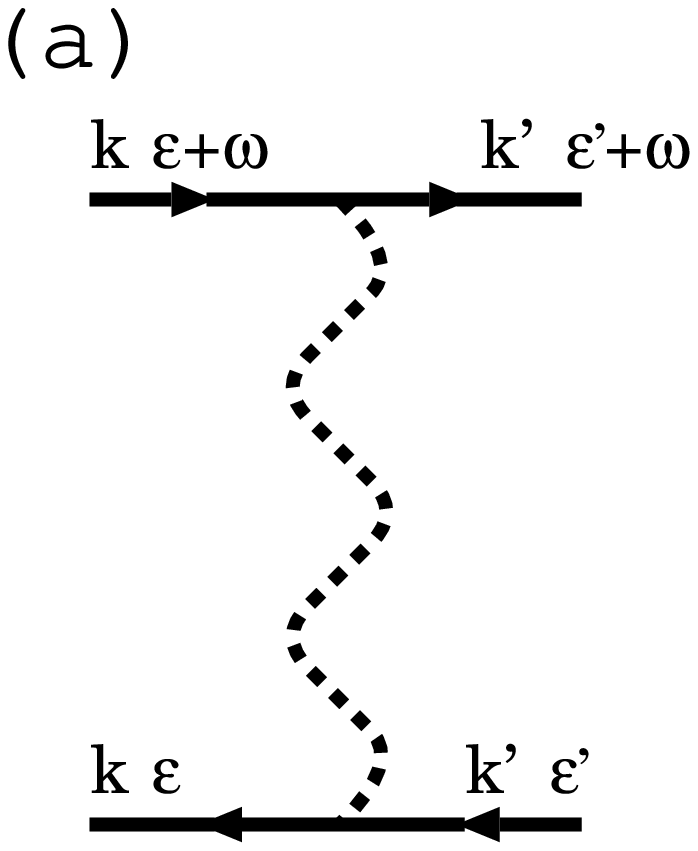}$$
\hspace{10mm}  
   \epsfysize=2.5cm
    $$\epsffile{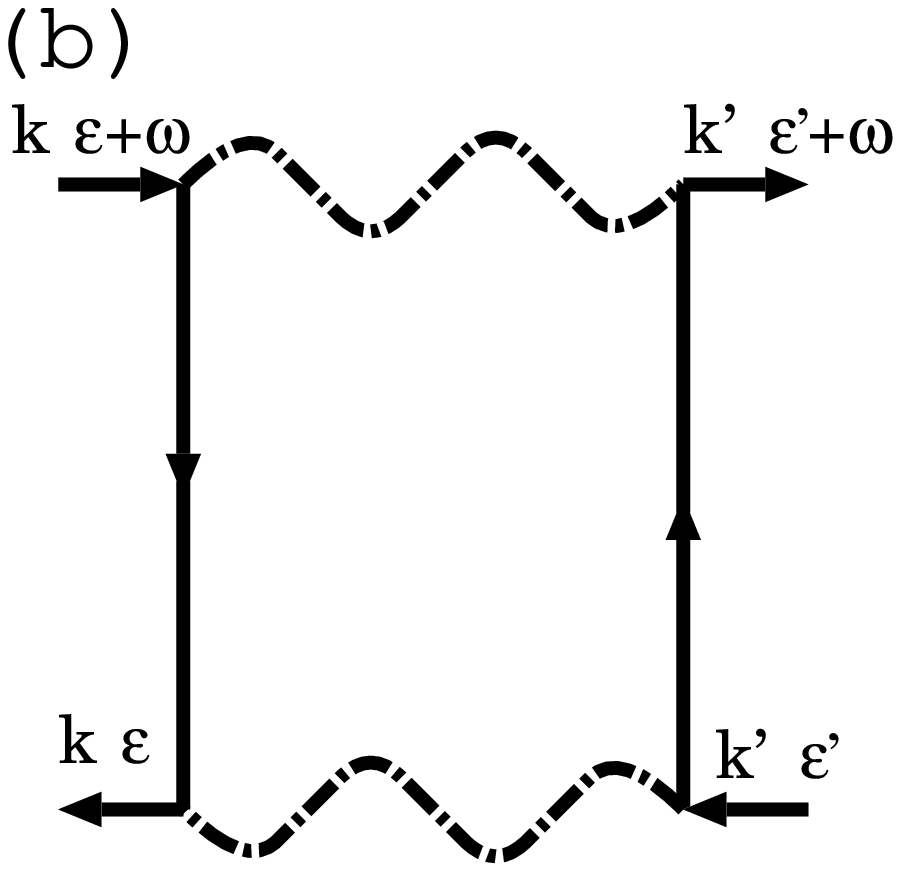}$$
\hspace{3mm}  
   \epsfysize=2.5cm
    $$\epsffile{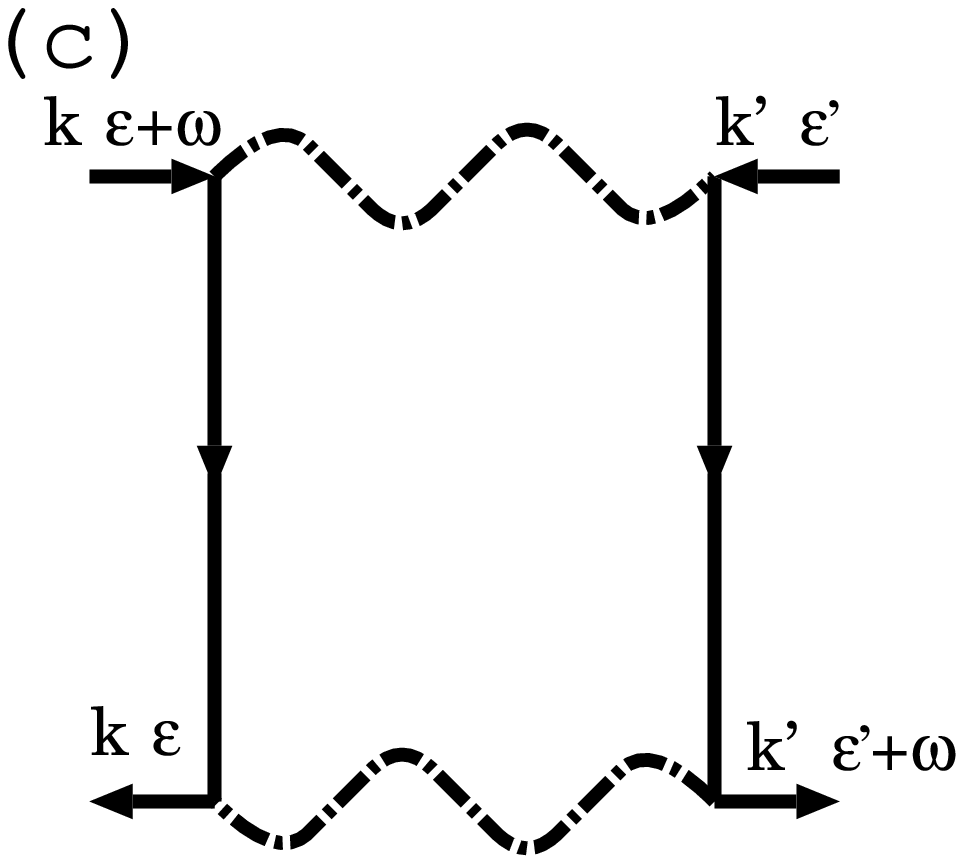}$$
    \caption{The four point vertex in the FLEX approximation. 
             (a) The SPMT term which is the dominant term. 
             The diagrams (b) and (c) are neglected in this paper. 
             }    
  \end{center}
\end{figure}

\begin{figure}[htbp]
  \begin{center}
   \epsfysize=6cm
    $$\epsffile{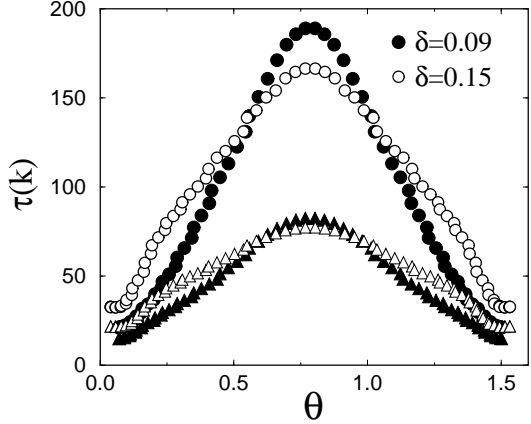}$$
    \caption{The momentum dependence of the lifetime $\tau(\k)$ on the Fermi 
             surface. 
             The horizontal axis shows the angle from {\it x}-direction, 
             $\theta={\rm Arctan}(k_{y}/k_{x})$. 
             The closed and open symbols correspond to
             the under-doped ($\delta=0.09$) and optimally-doped 
             ($\delta=0.15$) cases, respectively. 
             The circles and triangles are the results at $T=0.005$ and 
             $T=0.009$, respectively. 
             }    
  \end{center}
\end{figure}

 We point out three important properties which generate the unconventional 
transport in High-$T_{{\rm c}}$ cuprates. 
 The first is the momentum dependence of the quasi-particle's lifetime 
$\tau(\k)$. We show the typical results in Fig. 4 in which the horizontal 
axis is the angle from the $x$-axis $\theta={\rm Arctan}(k_{y}/k_{x})$. 
 It is clear that the lifetime is long in the diagonal direction 
$\theta=\pi/4$ and short around $\theta=0$ or 
$\pi/2$.~\cite{rf:hlubina,rf:stojkovic,rf:yanaseTR} 
 This property is caused by the momentum dependence of the AF 
spin fluctuations $\chi(\q,\Omega)$ which has the peak near the 
anti-ferromagnetic wave vector $\q \sim \Q=(\pi,\pi)$. 
 Since the quasi-particle's damping 
$\gamma(\k)=-{\rm Im}\s^{\rm R}(\k,0)$ depends on the 
low-energy density of state around $\kk = \k+\Q$, 
the lifetime is short near the magnetic Brillouin zone boundary on which 
$\e(\k)=\e(\k+\Q)$ and/or around the Van-Hove singularity at $\k=(\pi,0)$. 
This area is called \lq Hot spot'. 
 Therefore, the electric transport is practically carried 
by the quasi-particles at the Cold spot which is located near 
$\k=(\pi/2,\pi/2)$, namely $\theta=\pi/4$. 
 This momentum dependence plays an important role even in the pseudogap state.

 The second is the $T$-linear dependence of the damping rate 
$\gamma_{{\rm c}}= -{\rm Im} \s(\k_{{\rm c}},0)$ which causes the 
$T$-linear resistivity. Here, $\k_{{\rm c}}$ means the momentum 
at the Cold spot. 
 It have been pointed out that the $T^{2}$-resistivity is always 
obtained in the low temperature limit even at the quantum critical point 
unless the Fermi surface is perfectly nested.~\cite{rf:yanaseTR,rf:rosch} 
 However, the crossover temperature from $T^{2}$- to $T$-linear resistivity 
is sufficiently small because it decreases owing to the transformation of 
the Fermi surface.~\cite{rf:yanaseTR}

\begin{figure}[htbp]
  \begin{center}
   \epsfysize=6.5cm
    $$\epsffile{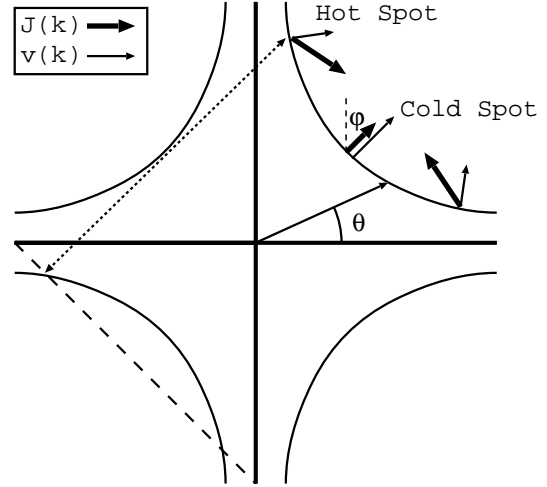}$$
    \caption{The schematic figure of the Fermi surface, the band velocity 
             $\vv(\k)$ and the current vertex $\J(\k)$. 
             The dashed line is the magnetic Brillouin zone boundary. 
             The thin and thick arrows correspond to $\vv(\k)$ and 
             $\J(\k)$, respectively. The dotted line shows the wave vector 
             $\Q=(\pi,\pi)$. The angles $\theta$ and $\varphi$ are used in  
             other figures. 
             }    
  \end{center}
\end{figure}

 The third is the vertex correction.~\cite{rf:kontani,rf:kanki} 
 The vertex correction is not so important for the resistivity. 
However, the correction from the SPMT term significantly enhances 
the Hall coefficient. 
 Although the momentum dependent lifetime also enhances the Hall 
coefficient,~\cite{rf:stojkovic,rf:yanaseTR} the vertex correction gives  
the dominant contribution to the enhancement.
 The temperature and doping dependence of the Hall coefficient is explained 
by this contribution. 
 Here, we explain the mechanism of the enhancement arising from the vertex 
correction.

 Since the AF spin fluctuations connect the current vertex 
$J(\k)$ with $J(\k+\Q)$, the vertex correction is especially important 
on the anti-ferromagnetic Brillouin zone boundary.  
 Near the Hot spot, the approximate solution of the Bethe-Salpeter equation 
is obtained by solving the simultaneous 
equations,~\cite{rf:kontani}
\begin{eqnarray}
   \J(\k) & = & \tilde{\vv}(\k) + \alpha(\k) \J(\k+\Q),  
\\
   \J(\k+\Q) & = & \tilde{\vv}(\k+\Q) + \alpha(\k+\Q) \J(\k).  
\end{eqnarray}
 Since we can assume $0< \alpha(\k+\Q) \cong \alpha(\k) < 1$ in this 
approximation, the current vertex is expressed as 
\begin{eqnarray}
  \J(\k) = \frac{\tilde{\vv}(\k)+\alpha(\k) \tilde{\vv}(\k+\Q)}
{1-\alpha(\k)^{2}}.     
\end{eqnarray}
 Since the velocity $\tilde{\vv}(\k+\Q)$ appears in the above expression, 
the directions of the velocity $\tilde{\vv}(\k)$ and 
the current vertex $\J(\k)$ are different. 
 Although the vertex correction is weaker at the Cold spot, 
the qualitatively same behavior as eq. (3.4) is expected. 
 The schematic figure of the band velocity $\vv(\k)$ and 
the current vertex $J(\k)$ is shown in Fig. 5.

\begin{figure}[htbp]
  \begin{center}
   \epsfysize=6cm
    $$\epsffile{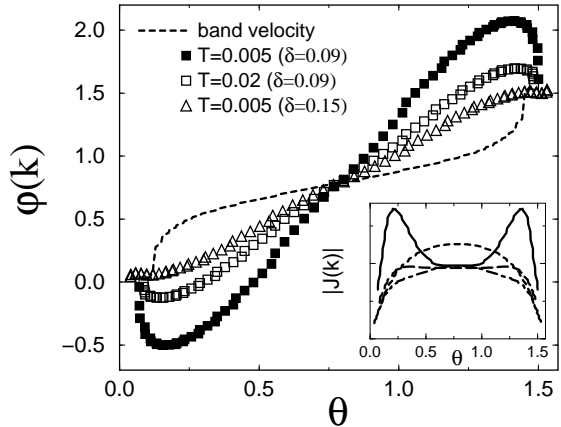}$$
    \caption{The angle of the current vertex 
             $\varphi(\k)={\rm Arctan}(J_{x}(\k)/J_{y}(\k))$ 
             on the Fermi surface. 
             The closed and open squares show the results 
             of the under-doped case ($\delta=0.09$) at 
             $T=0.005$ and $T=0.02$, respectively. 
             The triangles show the results at $\delta=0.15$ and $T=0.005$. 
             The dashed line is the angle of the band velocity 
             $\varphi(\k)={\rm Arctan}(v_{x}(\k)/v_{y}(\k))$ at $\delta=0.09$. 
             This angles $\varphi(\k)$ and $\theta$ are shown in Fig. 5.       
             The inset shows the absolute value $|\J(\k)|$. 
             The solid ($\delta=0.09$, $T=0.005$), 
             long dashed ($\delta=0.09$, $T=0.02$), 
             dash-dotted ($\delta=0.15$, $T=0.005$) lines are shown. 
             The dashed line shows the absolute value of the band velocity 
             $|\vv(\k)|$
             }    
  \end{center}
\end{figure}

 The transformation of the current vertex from the band velocity leads to 
the significant enhancement of the Hall coefficient. 
 This is because the Hall conductivity is approximately proportional to 
the gradient of the angle $\varphi(\k)={\rm Arctan}(J_{x}(\k)/J_{y}(\k))$ 
with respect to the momentum on the Fermi surface (See eq. (2.22)). 
 We show the obtained results of the angle $\varphi(\k)$ in Fig. 6. 
 The gradient at the Cold spot becomes steep as the 
spin fluctuations becomes strong. 
 Therefore, the Hall coefficient is temperature dependent   
in response to the temperature dependence of the spin fluctuations.

 The calculated resistivity $\rho$ and Hall coefficient $R_{{\rm H}}$ 
are shown in Fig. 7. 
 Fig. 7(a) shows the $T$-linear law of the resistivity in the sufficiently 
low temperature region. 
 It should be noticed that the absolute value of the current vertex $|J(\k)|$
is reduced from the band velocity $|\vv(\k)|$ at the Cold spot. 
 Therefore, the resistivity increases owing to the vertex correction. 
 However, the temperature dependence is not so affected since the increase is 
almost temperature independent (See the inset of Fig. 6). 
 Fig. 7(b) shows the enhancement of the Hall coefficient with decreasing the 
temperature. This temperature dependence is remarkable when the spin 
fluctuations are strong, namely in the under-doped region. 

 These results well explain the unconventional transport 
of High-$T_{{\rm c}}$ cuprates except for the pseudogap region. 
 The characteristic behaviors in the pseudogap state are explained 
by considering the SC fluctuations in the following sections.

\begin{figure}[htbp]
  \begin{center}
   \epsfysize=5.9cm
    $$\epsffile{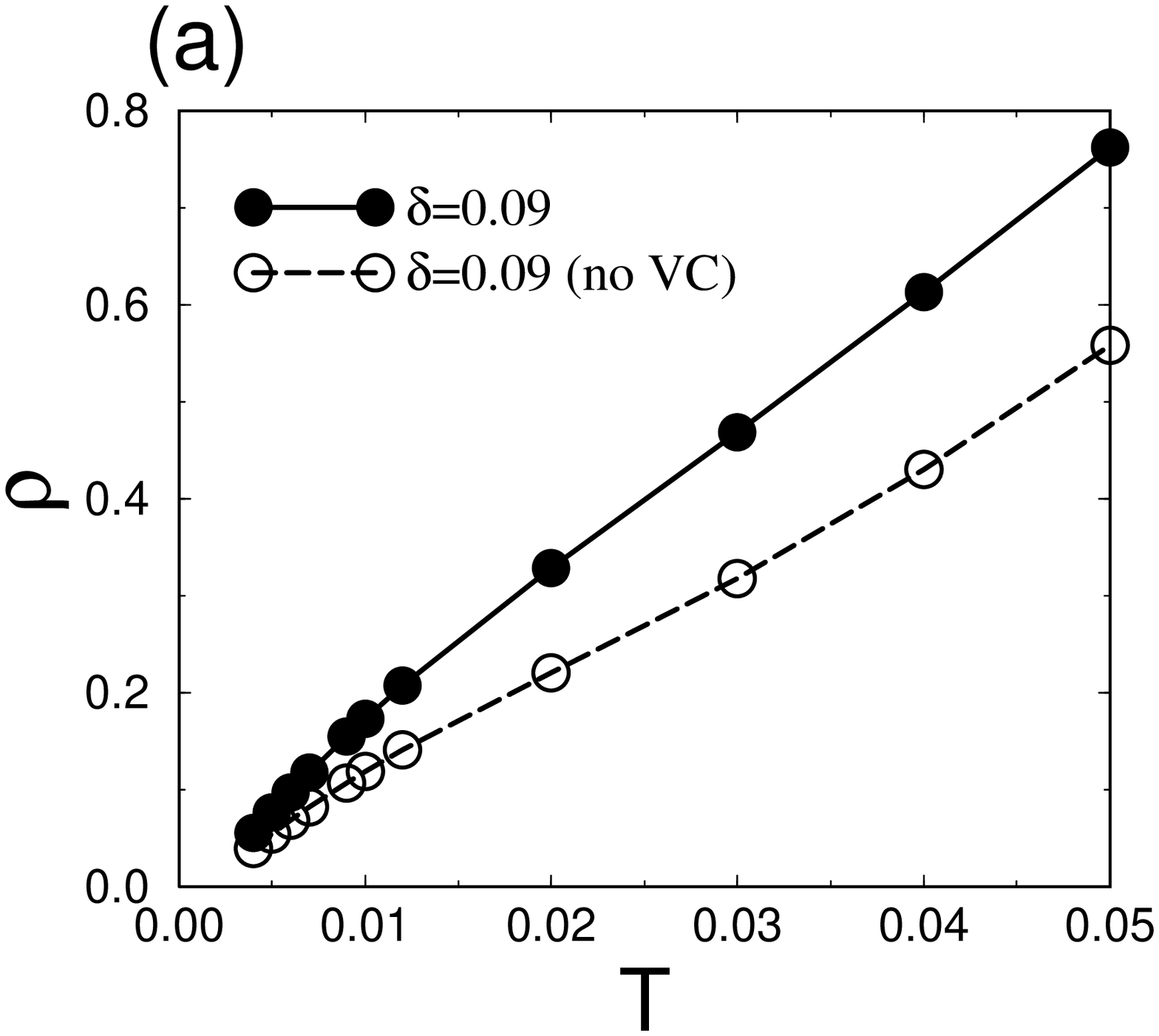}$$
\hspace{5mm}
   \epsfysize=5.8cm
    $$\epsffile{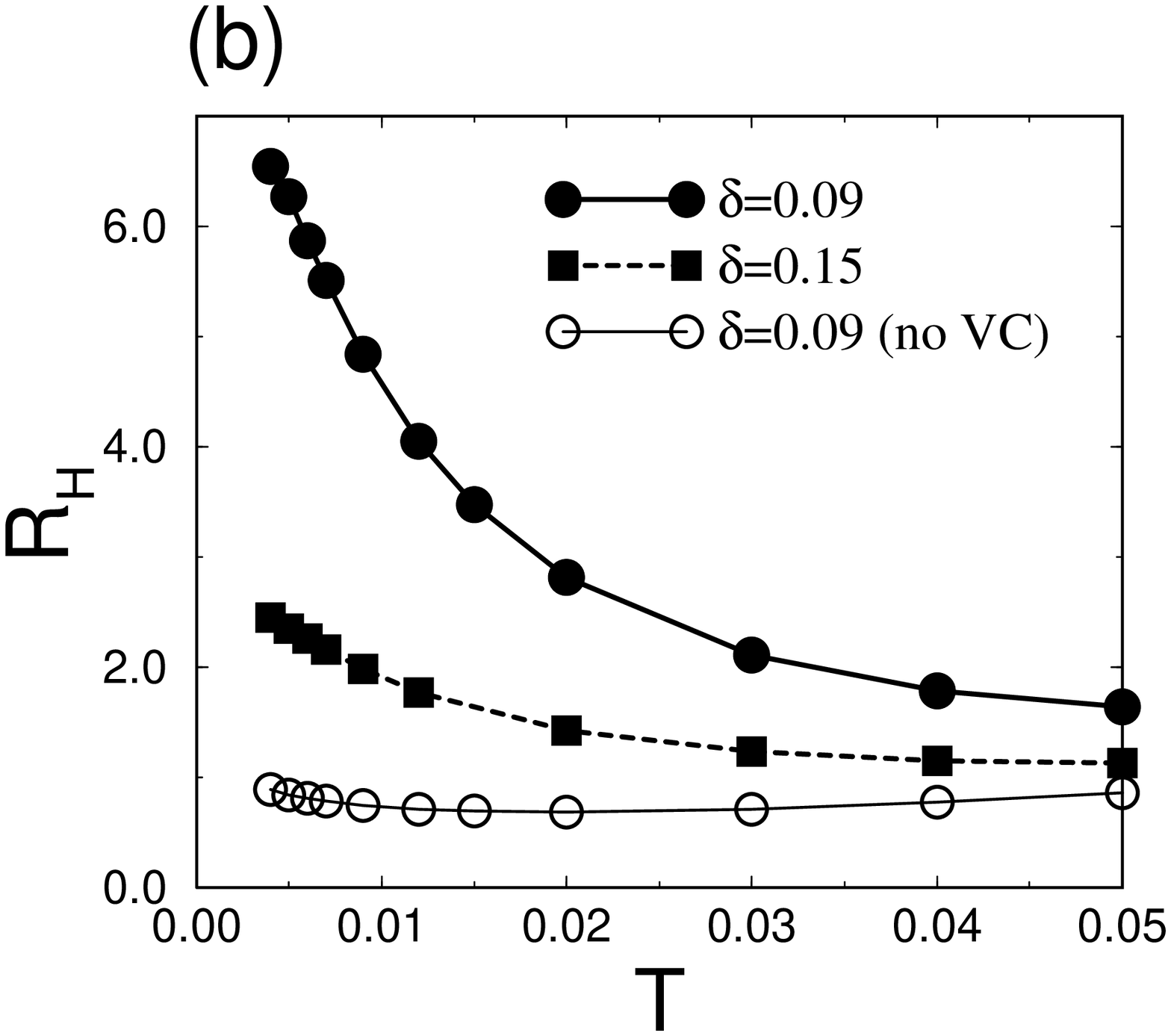}$$
    \caption{The results of the FLEX approximation at $U=1.6$. 
             (a)The resistivity 
             and (b) the Hall coefficient in under-doped 
             ($\delta=0.09$, circles) and optimally-doped
             ($\delta=0.15$, squares) cases. 
             The closed and open symbols show the results with and without 
             the vertex correction, respectively. 
             The shown temperature region is higher than 
             that in the following sections where the SC 
             fluctuations affect significantly. 
             }    
  \end{center}
\end{figure}

\section{Electric Transport in the Pseudogap State}

 In this section, we discuss the electric transport in the pseudogap state. 
 Because the SC fluctuations, spin-fluctuations and the single
particle properties are coupled to each other, there are many effects 
of the SC fluctuations. 
 First, we classify the effects of the SC fluctuations 
in the following way.

\begin{enumerate}
\item  The pseudogap in the single particle properties. 
\item  The feedback effects through the pseudogap in the AF spin 
       fluctuations.  
       This includes the decrease of the quasi-particle's damping 
       due to the spin fluctuations 
       $\gamma_{{\rm F}}(\k)=-{\rm Im}\s_{{\rm F}}^{\rm R}(\k,0)$.        
\item  The vertex corrections arising from the SC fluctuations. 
       The AL term is classified into these corrections. 
\end{enumerate}

 The following discussion and calculation show that the main effects 
are from (2). 
 Actually, we obtain a qualitatively same result 
by considering only the effects (2).  
 We estimate the effect (3) in the next section and show that the 
essential properties are obtained without the vertex correction 
arising from the SC fluctuations. Thus, in this section we show the results 
without the vertex corrections.

 We perform the SC-FLEX+T-matrix (SCFT) approximation in which the 
Green function and the spin susceptibility are obtained by solving 
eqs. (2.3)-(2.13) self-consistently. 
 In this section, the four point vertex is expressed by eq. (3.1) 
in which the spin susceptibility is suppressed by the pseudogap. 
 The repulsive interaction $U$ is chosen as $U=1.6$ or $U=2.0$, hereafter. 
 We show the results at $U=1.6$ for the comparison with 
the results of the FLEX approximation. 
Actually, the FLEX approximation is difficult in case of $U > 1.6$ and 
$\delta < 0.10$, because the system is too close to the 
anti-ferromagnetic instability. 
 The effects of the SC fluctuations are clarified by the comparison.  
 We can calculate at larger $U$ by considering the 
SC fluctuations, because the anti-ferromagnetic 
instability is suppressed by the pseudogap.~\cite{rf:yanaseFLEXPG}  
 Since the SC fluctuations become strong by increasing $U$, 
the pseudogap appears more clearly in case of $U=2.0$ than $U=1.6$.

 We first discuss the resistivity. 
 The effect (1) obviously reduces the longitudinal and transverse 
conductivities, because the extremely large damping is the origin of 
the pseudogap. 
 Therefore, our theory on the pseudogap seems to be incompatible with the 
downward deviation of the observed 
resistivity.~\cite{rf:ito,rf:takenaka,rf:odatransport,rf:mizuhashi}
 However, the increase of the resistivity by the effect (1) is not 
significant because 
the pseudogap occurs at the Hot spot which is not important for the 
transport phenomena. 
 On the other hand, the effect (2) which increases the conductivity 
gives the larger effect than (1), 
so that our results are consistent with the experiments. 
  When the spin fluctuations are suppressed by the pseudogap, 
the quasi-particle's damping by the spin fluctuations 
$\gamma_{{\rm F}}(\k)=-{\rm Im} \s_{{\rm F}}^{{\rm R}} (\k,0)$ is also 
suppressed. 
 Thus, the resistivity is decreased by the effect (2).

\begin{figure}[htbp]
  \begin{center}
   \epsfysize=6cm
    $$\epsffile{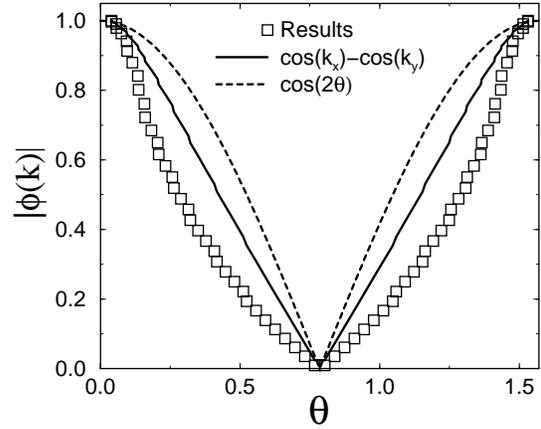}$$
    \caption{The momentum dependence of the wave function 
             $|\phi(\k)|=|\phi(\k,{\rm i} \pi T)|$ at $U=1.6$, 
             $\delta=0.09$ and $T=0.005$ (squares). 
              The results on the Fermi surface are shown. 
             The solid line shows the conventional momentum dependence 
             $|\phi(\k)| \propto |\cos(k_{x})-\cos(k_{y})|$. 
             The dashed line shows the line $|\phi(\k)| = |\cos(2 \theta)|$. 
             The results are normalized as the maximum value is unity
             $\phi(\k)|_{{\rm max}} = 1$. 
             Strictly speaking, we cannot define the Fermi surface at the 
             Hot spot because of the pseudogap. In this paper, 
             the Fermi surface is defined as the momentum at which the sign of 
             $\omega_{{\rm p}}$ changes. ($\omega_{{\rm p}}$ is the energy at 
             which the spectral weight $A(\k,\omega)$ has its maximum.) 
             The results are not affected by the definition, qualitatively. 
             }    
  \end{center}
\end{figure}

\begin{figure}[htbp]
  \begin{center}
   \epsfysize=6cm
    $$\epsffile{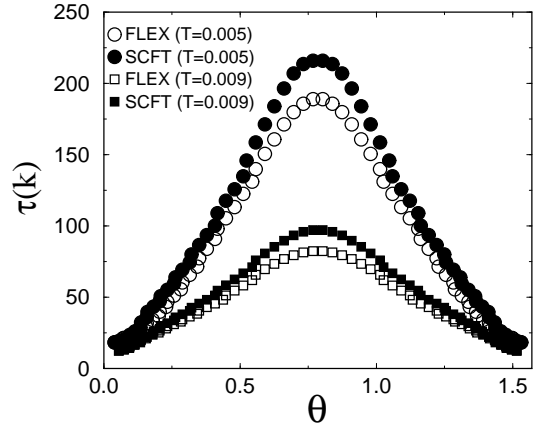}$$
    \caption{The momentum dependence of the lifetime $\tau(\k)$ on the Fermi 
             surface. 
             The open and closed symbols  
             show the results of the FLEX and the SCFT 
             approximations, respectively.  
             The circles and squares correspond to $T=0.005$ and  $T=0.009$, 
             respectively. 
             The parameters are chosen as $U=1.6$ and $\delta=0.09$. 
             }    
  \end{center}
\end{figure}

 We show the obtained momentum dependence of the wave function 
$\phi(\k)=\phi(\k,{\rm i} \pi T)$ in Fig. 8. 
 It is shown that the wave function at the Cold spot 
($\theta \sim \pi/4$) is smaller than that of the conventional 
$d$-wave form $\phi(\k) \propto \cos(k_{x})-\cos(k_{y})$ or 
$\phi(\k) = \cos(2 \theta)$. 
 This property becomes clearer as decreasing the hole-doping. 
 These results are consistent with the experimental results 
by the ARPES.~\cite{rf:mesot} 
 Because of the approximate relation 
$\gamma_{{\rm S}}(\k)=-{\rm Im} \s_{{\rm S}}^{{\rm R}} (\k,0) \propto |\phi(\k)|^{2} $,  
 the quasi-particle's damping induced by the SC fluctuations is 
further small at the Cold spot.

 In Fig. 9, we compare the quasi-particle's lifetime $\tau(\k)$ 
in the SCFT approximation and in the FLEX one. 
 It is clear that $\tau(\k)$ at the Cold spot ($\theta \sim \pi/4$) 
is longer in the SCFT approximation in which the effects (1) and (2) are 
taken into account. 
 Thus, the feedback effect (2) exceeds the effect (1) in the pseudogap 
state, and decreases the resistivity.

 The other important character is that the increase of the 
$\tau(\k)$ is slight, while the $1/T_{1}T$ is remarkably reduced 
(See Fig. 2(b)). 
 This is because the Cold spot is not so sensitive to the feedback effect, 
while the Hot spot is significantly affected. 
 As is shown in Ref. 36, the suppression of $\chi(\q,\omega)$ by the pseudogap 
is strong at $\q=\Q$, while the suppression of the incommensurate component is 
comparatively small. 
 Since the quasi-particles at the Cold spot are not directly scattered by 
the spin fluctuations at $\q=\Q$, the feedback effect is relatively small. 
 Therefore, the resistivity only slightly decreases 
owing to the SC fluctuations. 
 This is not caused by the competition between the effects (1) and (2), 
however caused by the slightness of the effect (2).

\begin{figure}[htbp]
  \begin{center}
   \epsfysize=6.2cm
    $$\epsffile{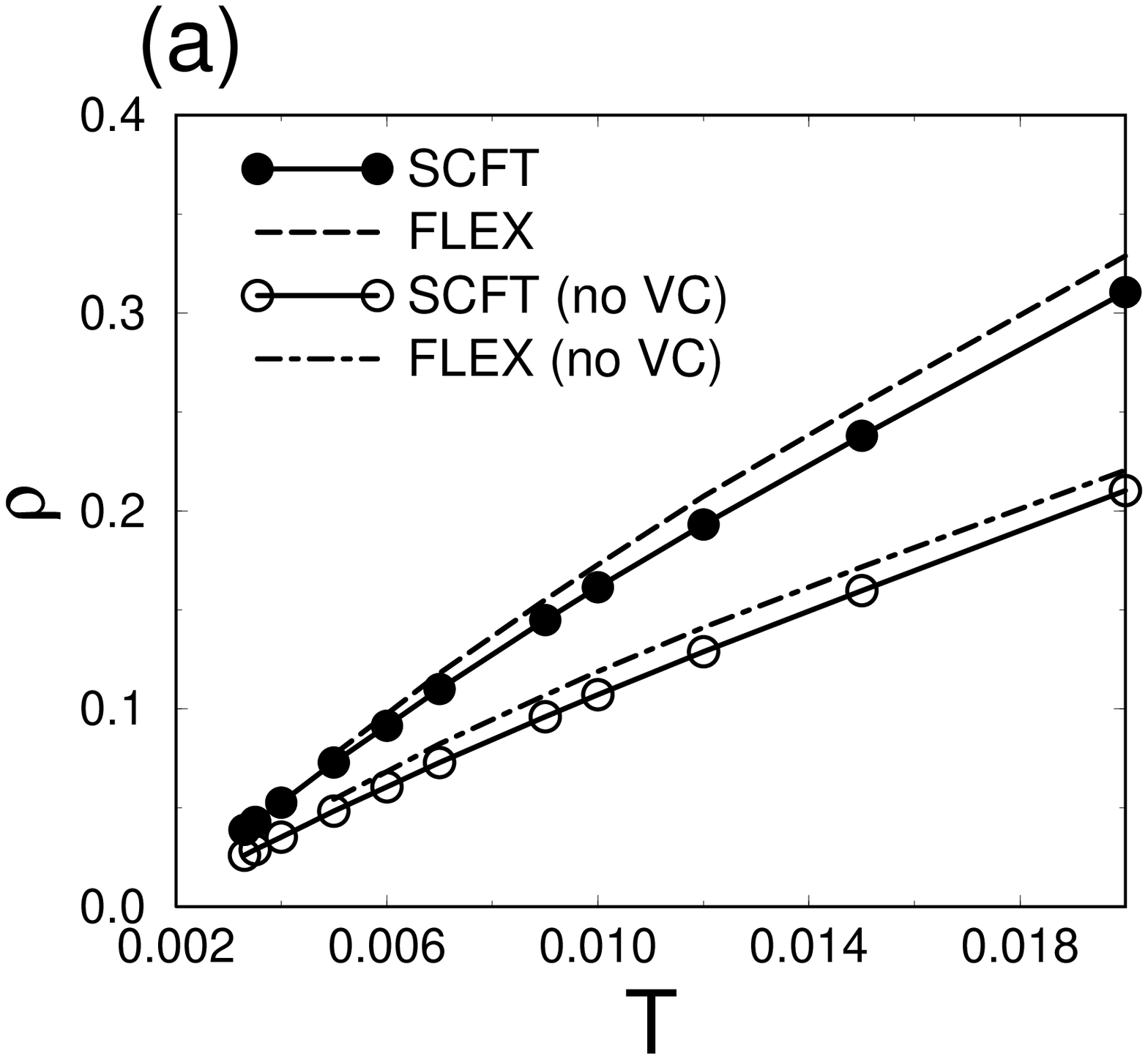}$$
\hspace{5mm}
   \epsfysize=6.2cm
    $$\epsffile{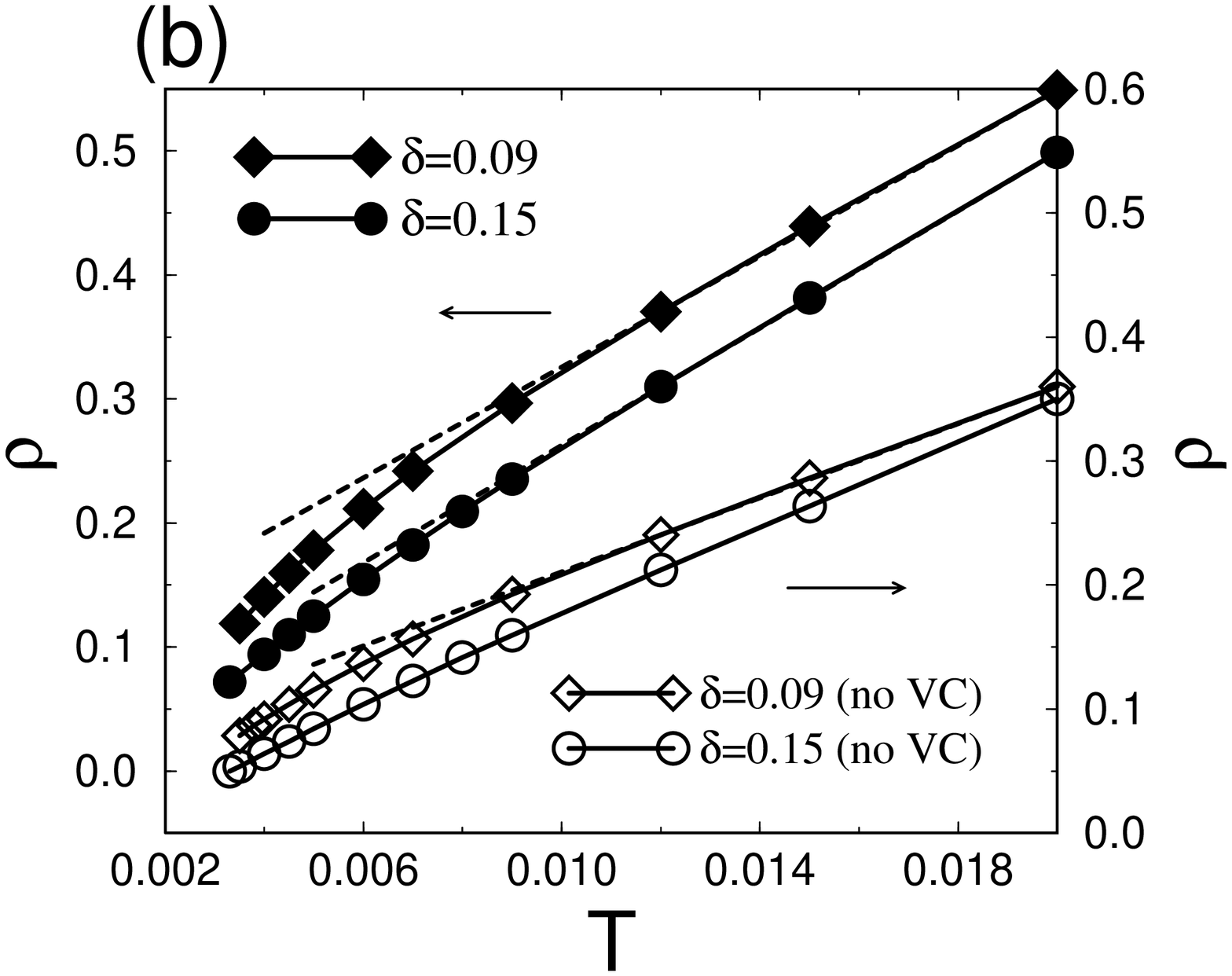}$$
    \caption{The temperature dependence of the resistivity $\rho$. 
             (a) The results at $U=1.6$ and $\delta=0.09$. 
             The closed and open circles 
             show the results with and without the SPMT term, 
             respectively. 
             The long-dashed and dash-dotted lines show the results of the  
             FLEX approximation. 
             (b) The results at $U=2.0$. The squares and circles correspond to 
             under-doped ($\delta=0.09$) and optimally-doped ($\delta=0.15$)
             cases, respectively. 
             The closed and open symbols show the results with and 
             without the SPMT term, respectively. 
             The vertical axis of the closed symbols is shifted for  
             eyes. The additional dashed lines are also shown for eyes. 
             }    
  \end{center}
\end{figure}

 The obtained results for the resistivity are shown in Fig. 10. 
The comparison between the SCFT and FLEX approximations at 
$U=1.6$ (Fig. 10(a)) confirms the above discussion which indicates that 
the effect (2) exceeds the effect (1). 
 Since the downward deviation occurs even in the FLEX approximation, 
this deviation does not necessarily mean the onset of the pseudogap. 
 Indeed, the downward deviation is a feature in the crossover region 
from the $T$-linear law to the $T^{2}$-law of the resistivity. 
 Anyway, the feedback effect also contributes to the downward deviation.

 In case of $U=2.0$ the resistivity shows the downward deviation more clearly 
(Fig. 10(b)). 
 This feature becomes more remarkable as decreasing the hole-doping, 
which is consistent with the experimental results including the doping 
dependence. 
 It is notable that downward deviation becomes more distinct 
by considering the vertex correction. 
 This means that the feedback effect through the SPMT term is also 
important. 
 Anyway, an important result is that the response of the resistivity 
to the pseudogap is remarkably weak compared with the magnetic properties
and the single particle properties. 
 Thus, our theory is consistent with the experimental results in detail.~\cite{rf:ito,rf:takenaka,rf:odatransport,rf:mizuhashi} 

 It is worthwhile to comment that the approximate relation 
$\rho \propto T^{2} \chi(\Q,0)$ or  $\rho \propto T^{2}/T_{1}T$, which has been
used in the spin fluctuation theory, 
considerably over-estimates the feedback effect. 
 The appropriate results are obtained by fully considering the  
momentum dependence.

 Next, we discuss the Hall coefficient. 
 The Hall coefficient is enhanced by the effect (1) 
because the momentum dependence of the lifetime becomes strong. 
 Actually, this enhancement by the SC fluctuations was 
used to explain the enhancement of the Hall coefficient.~\cite{rf:ioffe} 
 However, we find that this effect is much smaller than the feedback 
effect (2). 
 The behavior of the Hall coefficient is mainly determined by the vertex 
correction arising from the SPMT term, since this term remarkably enhances 
the Hall coefficient. 
 The Hall coefficient are obviously reduced by the feedback effect (2). 
 This feedback effect is much more significant than that on the resistivity. 
 The momentum dependence of the angle 
$\varphi(\k)={\rm Arctan}(J_{x}(\k)/J_{y}(\k))$ is shown in Fig. 11(a). 
 It is shown that the gradient at the Cold spot ($\theta \sim \pi/4$) 
is remarkably reduced by the SC fluctuations, 
however still larger than the band velocity. 
 Fig. 11(b) show the absolute value of the current vertex $|J(\k)|$.  
 It is clear that the feedback effect on $\varphi(\k)$ at the Cold spot 
is much stronger than that on $|J(\k)|$. 
 We can understand from Fig. 11(b) that the feedback effect through the SPMT 
term is significant at the Hot spot where the pseudogap occurs.

\begin{figure}[htbp]
  \begin{center}
   \epsfysize=6cm
    $$\epsffile{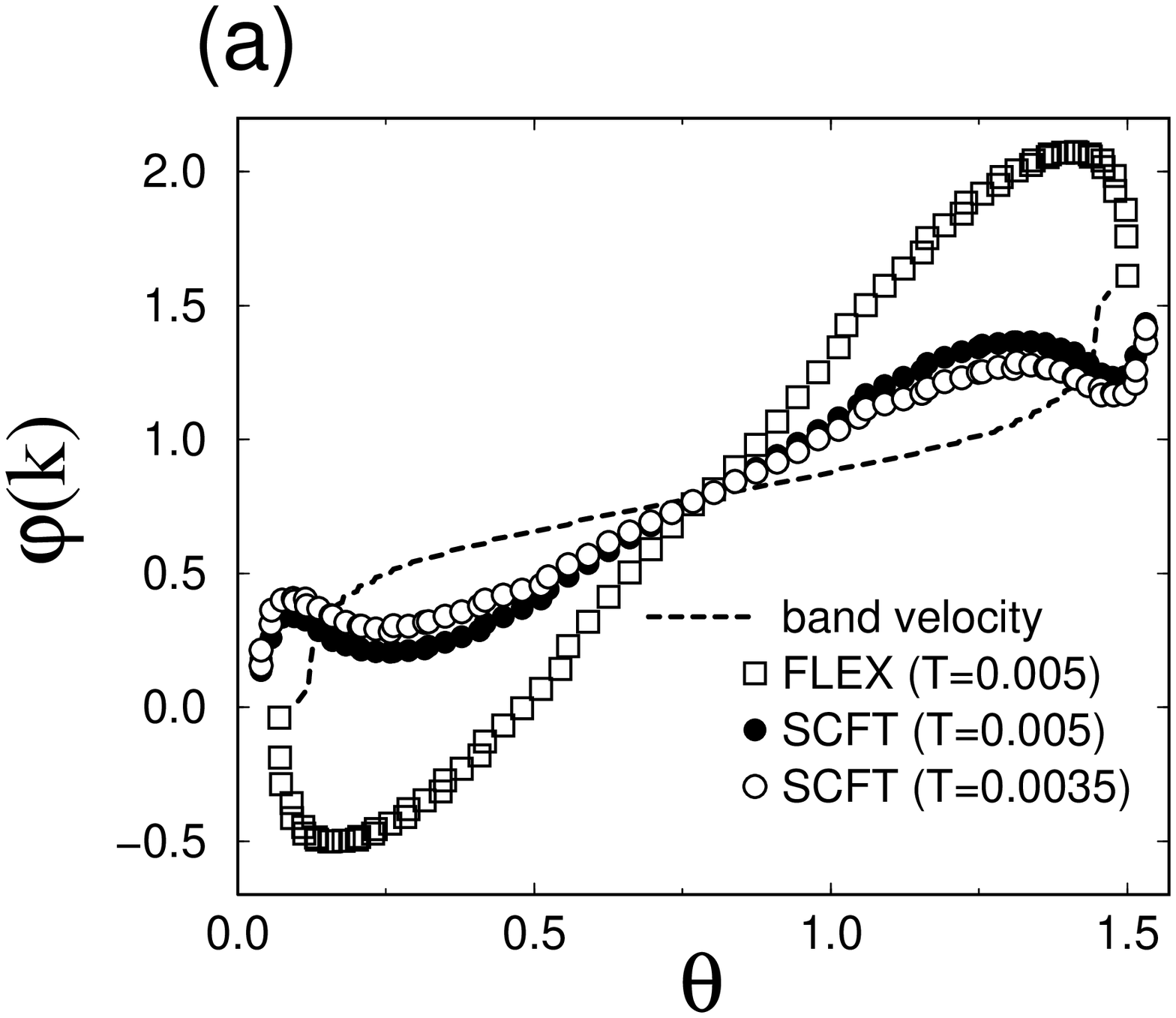}$$
   \epsfysize=6cm
    $$\epsffile{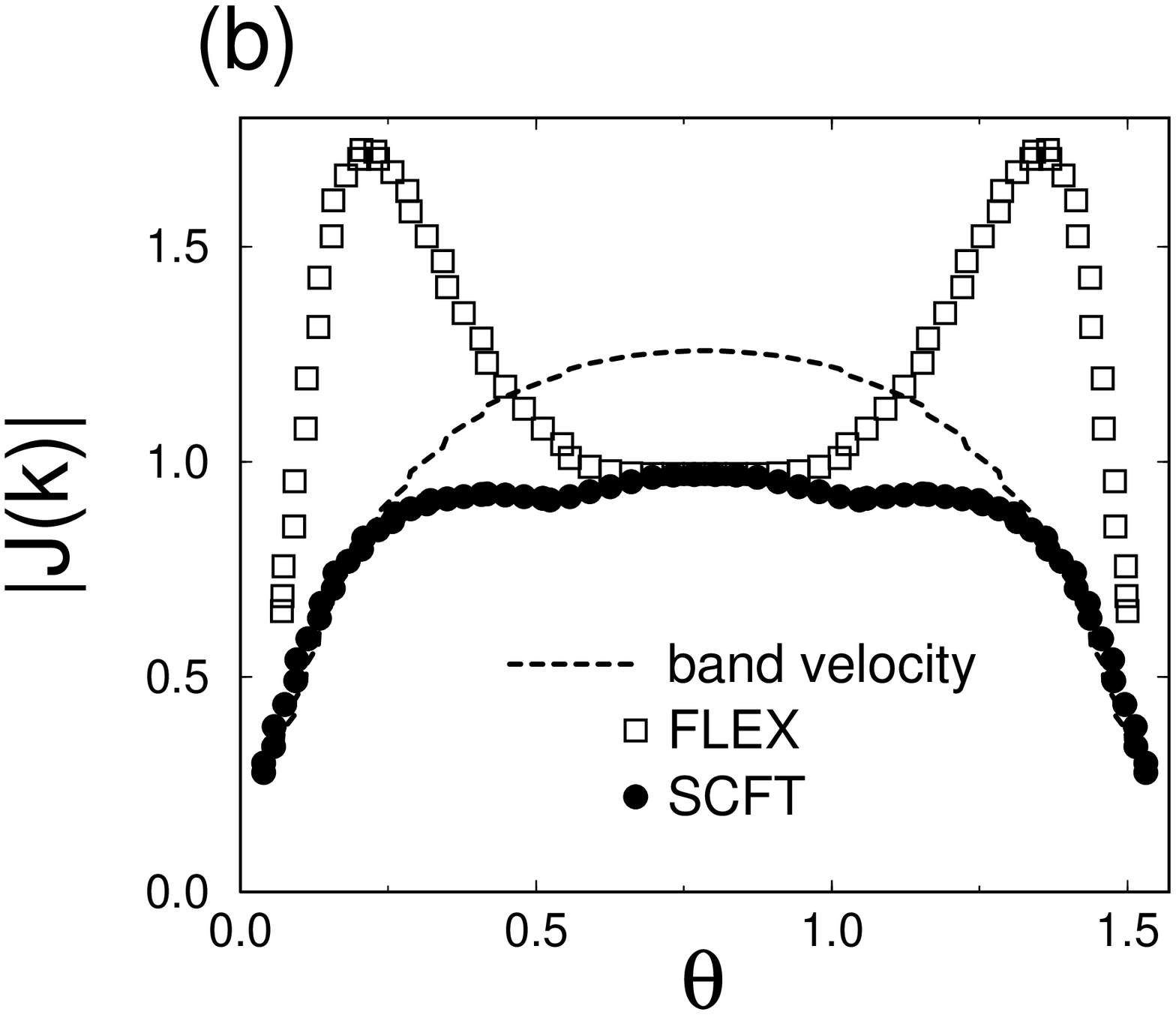}$$
    \caption{(a) The angle of the current vertex 
             $\varphi(\k)={\rm Arctan}(J_{x}(\k)/J_{y}(\k))$ 
             on the Fermi surface. 
             The open and closed circles show the results of the 
             SCFT approximation at $T=0.0035$ and $T=0.005$, 
             respectively.  
             The squares show the result of the FLEX approximation at 
             $T=0.005$. 
             The dashed line shows the angle of the band velocity 
             $\varphi(\k)={\rm Arctan}(v_{x}(\k)/v_{y}(\k))$. 
             (b) The absolute value of the current vertex $|\J(\k)|$  
             in the FLEX (squares) and SCFT (circles) 
             approximations at $T=0.005$. 
             The dashed line shows the absolute value of the band velocity 
             $|\vv(\k)|$. 
             The parameters are chosen as $U=1.6$ and $\delta=0.09$.  
             }    
  \end{center}
\end{figure}

 The calculated results of the Hall coefficient are shown in Fig. 12 which  
clearly confirm the above discussions. 
 The Hall coefficient at $U=1.6$ is remarkably reduced from the result of the 
FLEX approximation. The result shows the peak near $T=0.005$ 
and decreases with decreasing the temperature (Fig. 12(a)). 
 The decrease is more distinct in case of $U=2.0$ (Fig. 12(b)). 
 The decrease of the Hall coefficient in the pseudogap state becomes small 
with increasing the hole-doping. 
 These results qualitatively explain the experimental results in the pseudogap 
state,~\cite{rf:ito,rf:satoM,rf:haris,rf:mizuhashi,rf:xiong,rf:ong} 
including the doping dependence. 
 If we neglect the SPMT term, the Hall coefficient shows only the 
slight increase with  decreasing the temperature 
(See the inset of Fig. 12(a)). 
 In this case, the absolute value is much smaller than the experimental 
results. 
 Thus, the SPMT term plays an essential role for explaining 
the Hall coefficient even in the pseudogap state.

\begin{figure}[htbp]
  \begin{center}
   \epsfysize=6cm
    $$\epsffile{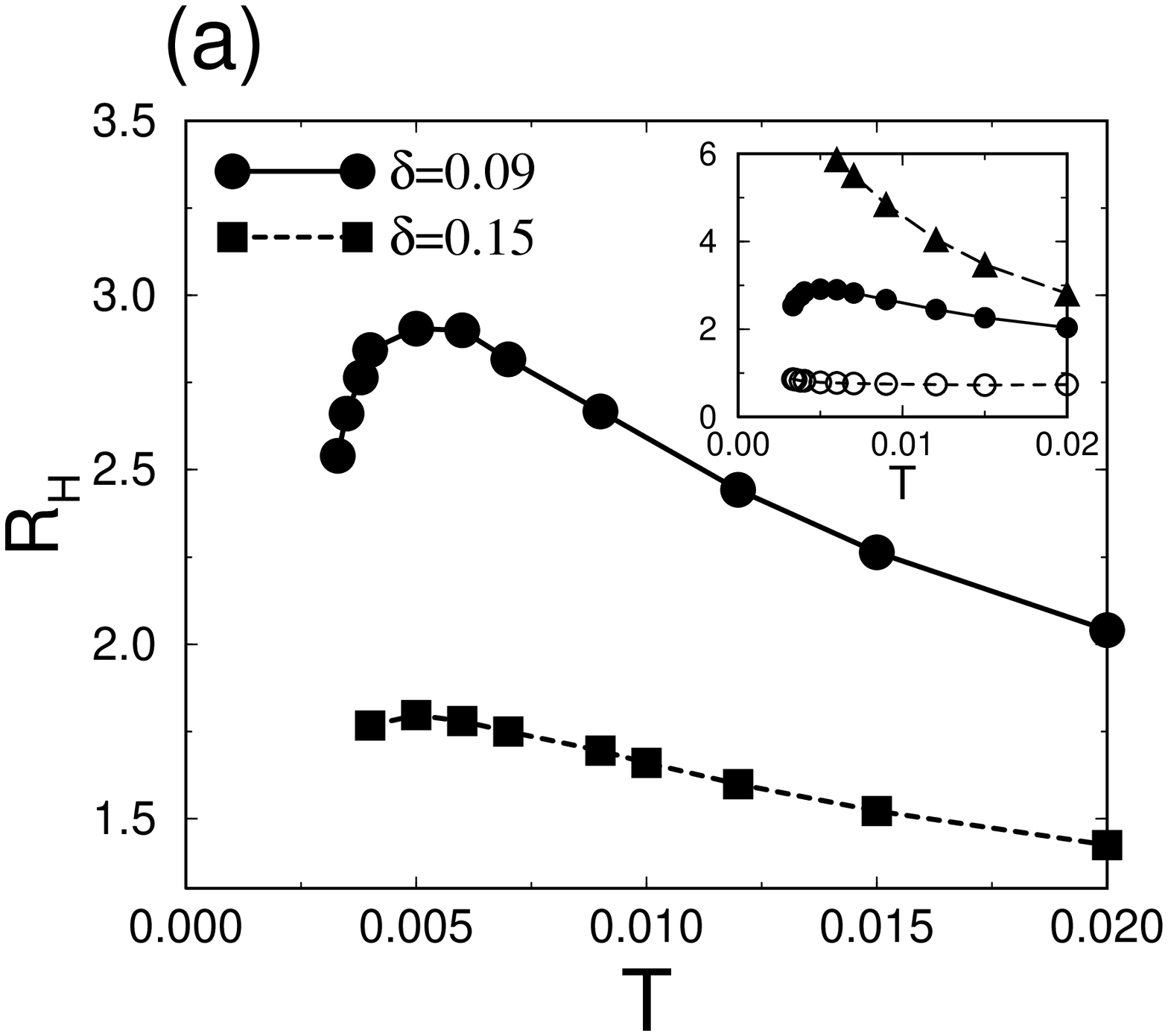}$$
   \epsfysize=6cm
    $$\epsffile{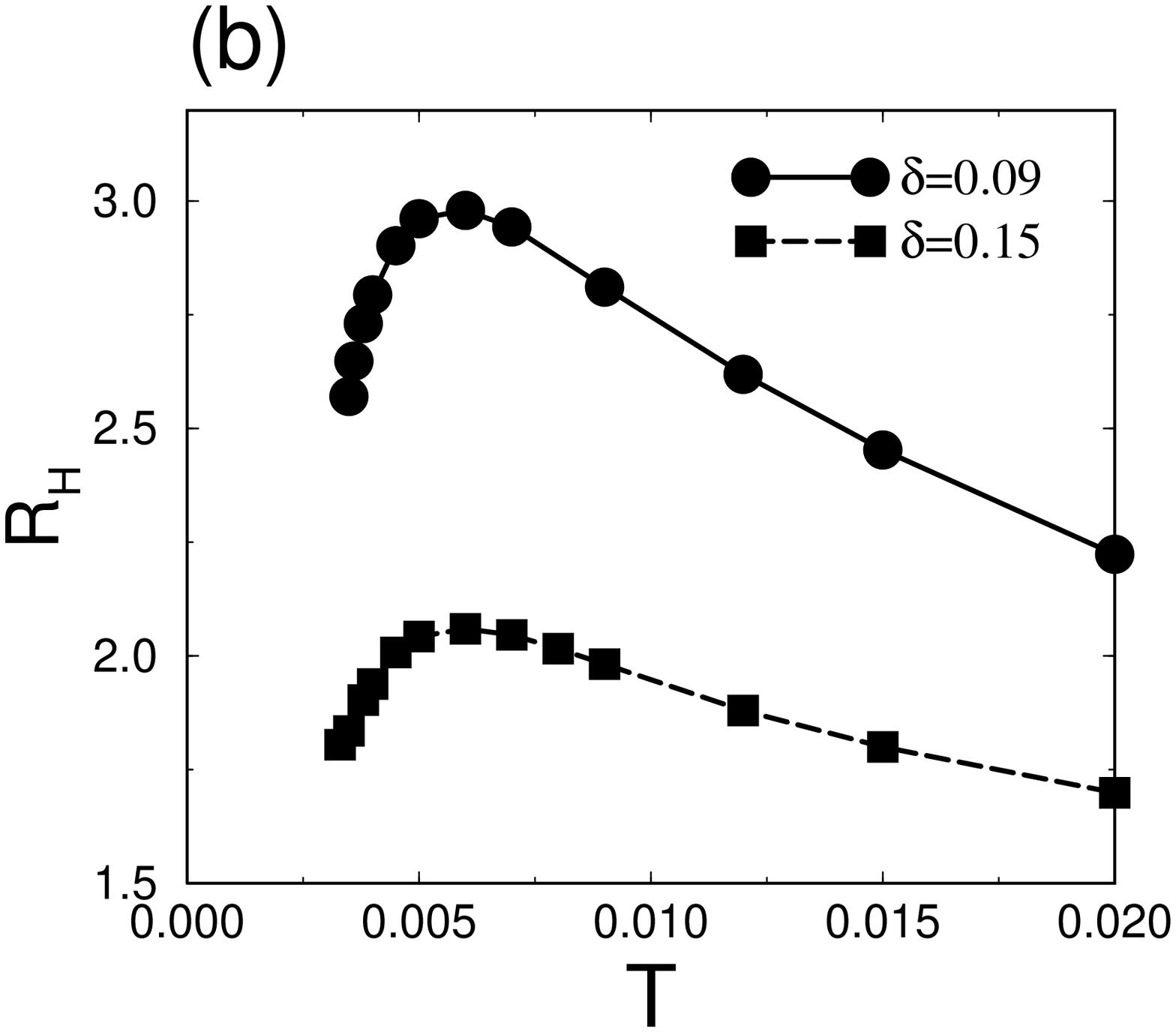}$$
    \caption{The temperature dependence of the Hall coefficient $R_{{\rm H}}$.
             (a) The results at $U=1.6$ obtained 
             by the SCFT approximation 
             at $\delta=0.09$ (circles) and $\delta=0.15$ (squares). 
             The inset shows the comparison with  
             the FLEX approximation (triangles) and the SCFT 
             approximation without the SPMT term (open circles) at 
             $\delta=0.09$. 
             (b) The results at $U=2.0$. The circles and squares correspond to 
             $\delta=0.09$ and $\delta=0.15$, respectively. 
             }    
  \end{center} 
\end{figure}

 Thus, the behaviors of the transport coefficients 
in the pseudogap state are explained by considering the spin fluctuations and 
the SC fluctuations simultaneously. 
 It is confirmed that these properties are mainly caused by the feedback 
effect through the pseudogap of the spin fluctuations. 
 The feedback effect is sufficiently small for the resistivity, however 
is rather strong for the Hall coefficient. 
 This difference reflects the importance of the SPMT term for the quantities. 
 Thus, the different responses of the respective quantities are naturally 
explained in our theory.

\section{Vertex Corrections by the Superconducting Fluctuations}

 In this section, the vertex corrections arising from the 
SC fluctuations are discussed. 
 Since the SC fluctuations also give rise to the temperature 
dependence of the four point vertex, it is worthwhile to investigate 
the effects of the SC fluctuations through the vertex correction. 
 It is difficult to satisfy the Ward identity including the SC 
fluctuations, because the SCFT approximation is not the 
conserving approximation. 
 However, this is not crucial in terms of the momentum conservation law 
since the conservation law has already been broken in the lattice systems. 
 Here, we estimate the vertex corrections in a perturbative way with respect 
to the SC fluctuations 
in order to get a qualitative knowledge about their effects.

 In addition to that, the SC fluctuations have a divergent contribution 
which is not included in the \eli formalism. 
 Therefore, we estimate the AL term which is the most 
possible term among them. 
 A main conclusion of this section is that the above corrections do not give 
a qualitative modification of the results in \S4.  
 That is, the discussions in \S4 are robust in a qualitative sense.

\begin{figure}[htbp]
  \begin{center}
   \epsfysize=2.7cm
    $$\epsffile{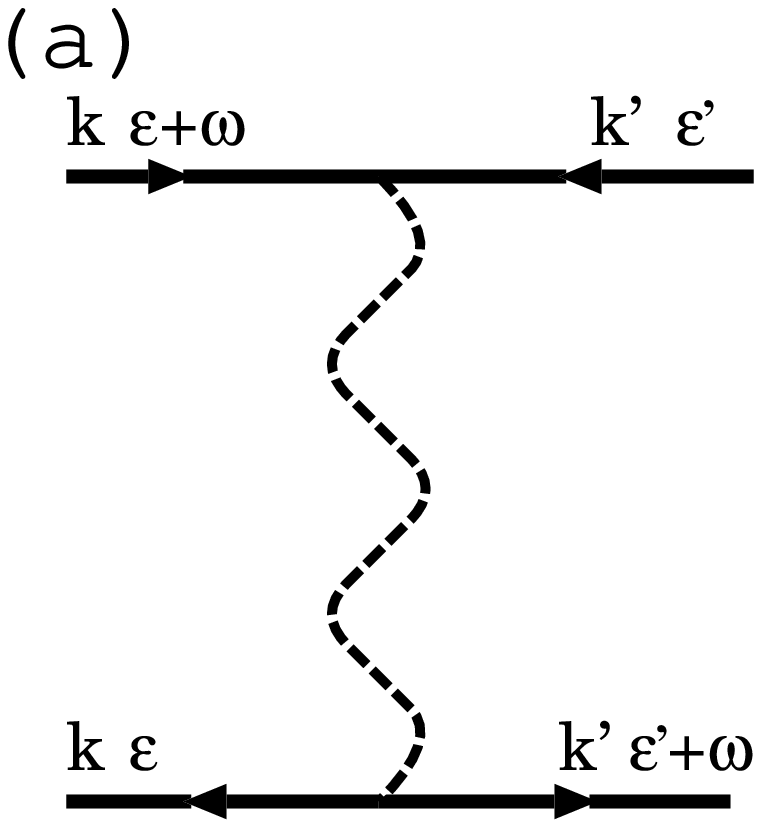}$$
\hspace{3mm}  
   \epsfysize=2.2cm
    $$\epsffile{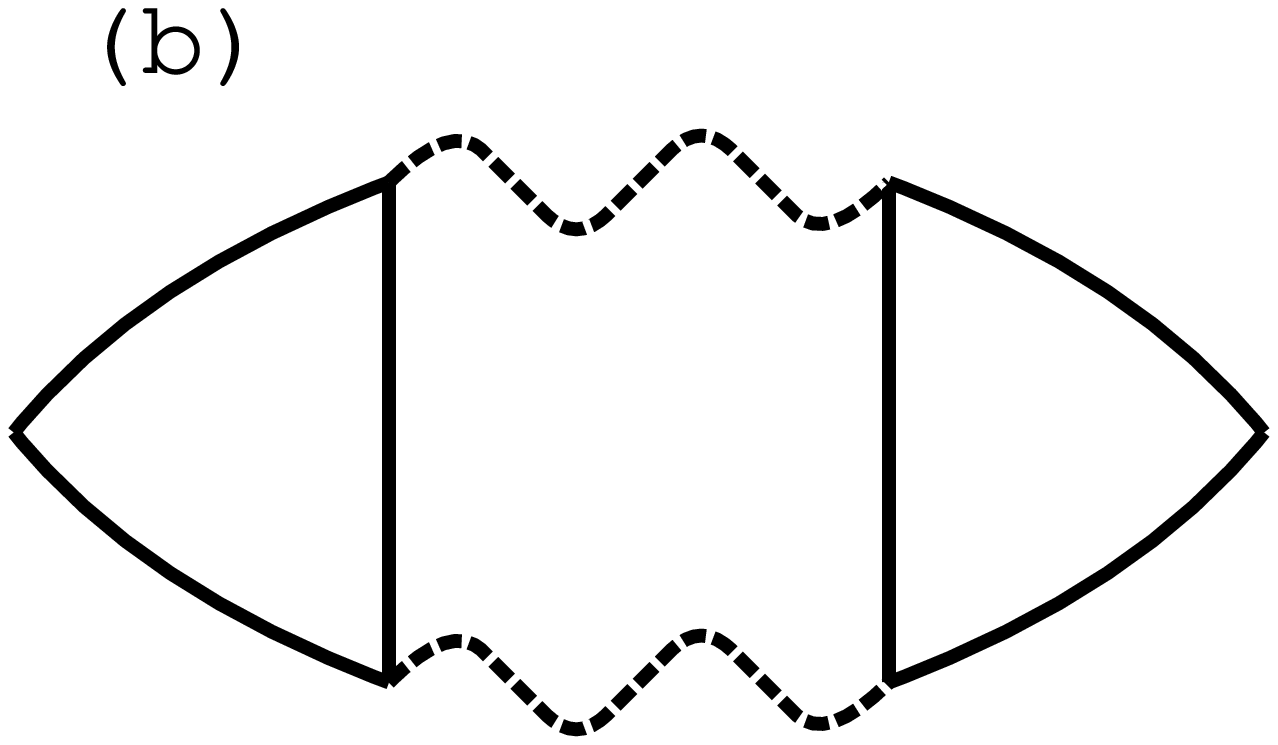}$$
    \caption{(a) The four point vertex corresponding to eq. (5.1). 
             (b) The corresponding Feynman diagram of the AL term. 
             The solid line and the long-dashed wavy line represent the 
             propagator of the single particles and the SC 
             fluctuations, respectively. 
             }    
  \end{center}
\end{figure}

 First, we consider the vertex correction within the lowest order about 
$1/\tau(\k)$. Here, the parameter $1/\tau(\k)$ is interpreted as that at 
the Cold spot $1/\tau_{{\rm c}}=1/\tau(\k_{{\rm c}})$. 
 We estimate only the lowest order term with respect to the T-matrix 
(Fig. 13 (a)). 
 The term is written as the superconducting fluctuation Maki-Thompson (SCMT) 
term in this paper. 
 Afterward, the higher order term is partly estimated and 
is confirmed to be negligible in the main part of the pseudogap state. 
 The SCMT term has a similar, but not the same contribution as the 
Maki-Thompson term.~\cite{rf:MT} 
 Because we treat this term by using the \eli formalism, 
the vertex correction is calculated iteratively, however  
the higher order correction with respect to $1/\tau(\k)$ is not included.

 The four point vertex function of the SCMT term is obtained by using the  
procedure in \S2, 
\begin{eqnarray}
  \Im_{22}^{{\rm S}}(\k,\e:\kk,\ee) &=& 
  2 {\rm i} |\phi(\k,\e)|^{2} 
  ({\rm cth}(\frac{\e+\ee}{2 T}) - {\rm tanh}(\frac{\ee}{2 T}))
\nonumber \\
&\times&
  {\rm Im} t^{{\rm R}} (\k+\kk,\e+\ee). 
\end{eqnarray}
 The total four point vertex is given by the summation  
of eq. (3.1) and eq. (5.1), $\Im_{22}=\Im_{22}^{{\rm F}}+\Im_{22}^{{\rm S}}$. 
 It should be noticed that the vertex correction by the SCMT term disappears 
in the diagonal direction $\theta=\pi/4$ because of the $d$-wave form factor
(See Fig. 8). 
 Therefore, it is expected that the SCMT term does not play an essential role. 
 The characteristic momentum dependence (Hot spot and Cold spot) 
plays an important role also in this stage.

 The SCMT term couples the current vertex $J(\k)$ to $J(-\k)=-J(\k)$. 
 Then, the similar estimation to eqs.(3.2-4) reveals the following 
approximate relation, 
 \begin{eqnarray}
   \J(\k) & = & \tilde{\vv}(\k) + \alpha(\k) \J(\k+\Q) + \beta(\k) \J(\k), 
\\
   \J(\k+\Q) & = & \tilde{\vv}(\k+\Q) + \alpha(\k+\Q) \J(\k) 
\nonumber \\
& & + \beta(\k+\Q) \J(\k+\Q),   
\\
  \J(\k) & = & \frac{(1-\beta(\k)) 
           \tilde{\vv}(\k)+\alpha(\k) \tilde{\vv}(\k+\Q)}
           {(1-\beta(\k))^{2} -\alpha(\k)^{2}}.     
\end{eqnarray}
 The factor $\beta(\k)$ arises from the SCMT term, and 
satisfies the relation $0 \leq \beta(\k) < 1-\alpha(\k)$. 
 A simple estimation reveals that the current vertex 
increases by the SCMT term {\it i.e.}, the SCMT term enhances the 
longitudinal conductivity. 

 Since the coefficient of $\tilde{\vv}(\k)$ in eq. (5.4) relatively decreases  
by the factor $1-\beta(\k)$, the gradient of the angle $\varphi(\k)$ 
becomes steep by the SCMT term. 
 However, this effect is not so strong at the Cold spot 
because the factor $\beta(\k)=0$ at $\theta=\pi/4$. 
 These discussions are confirmed by showing the calculated results 
of the angle $\varphi(\k)$ in Fig. 14 (a). 
 As a result, the Hall coefficient is enhanced by the SCMT term, however 
the enhancement is not significant. 
 It is notable that the enhancement of the Hall coefficient does not 
occur without the SPMT term. 
 In other wards, the SCMT term indirectly enhances the Hall coefficient 
through the effect of the SPMT term.

\begin{figure}[htbp]
  \begin{center}
   \epsfysize=6cm
    $$\epsffile{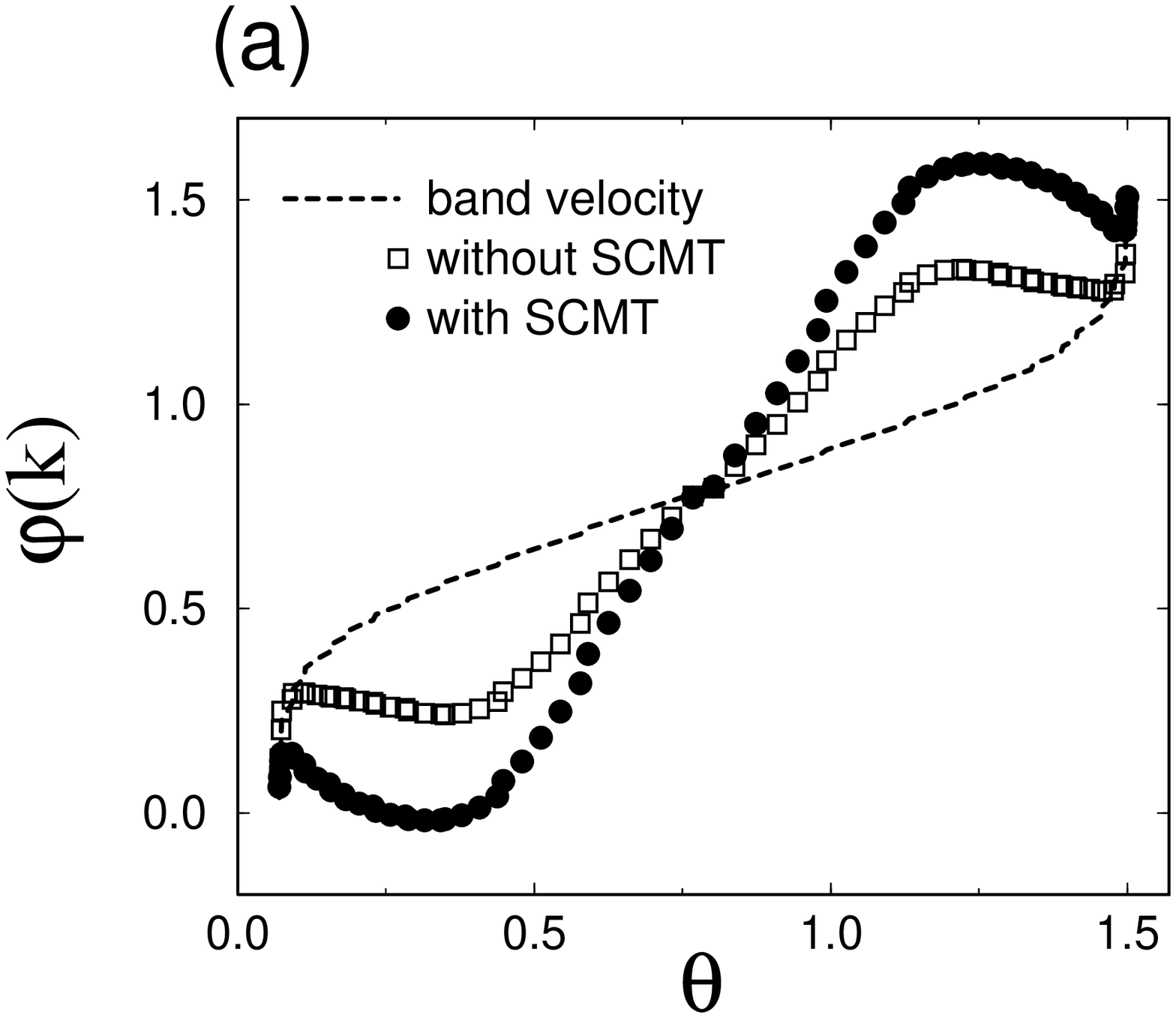}$$
   \epsfysize=6cm
    $$\epsffile{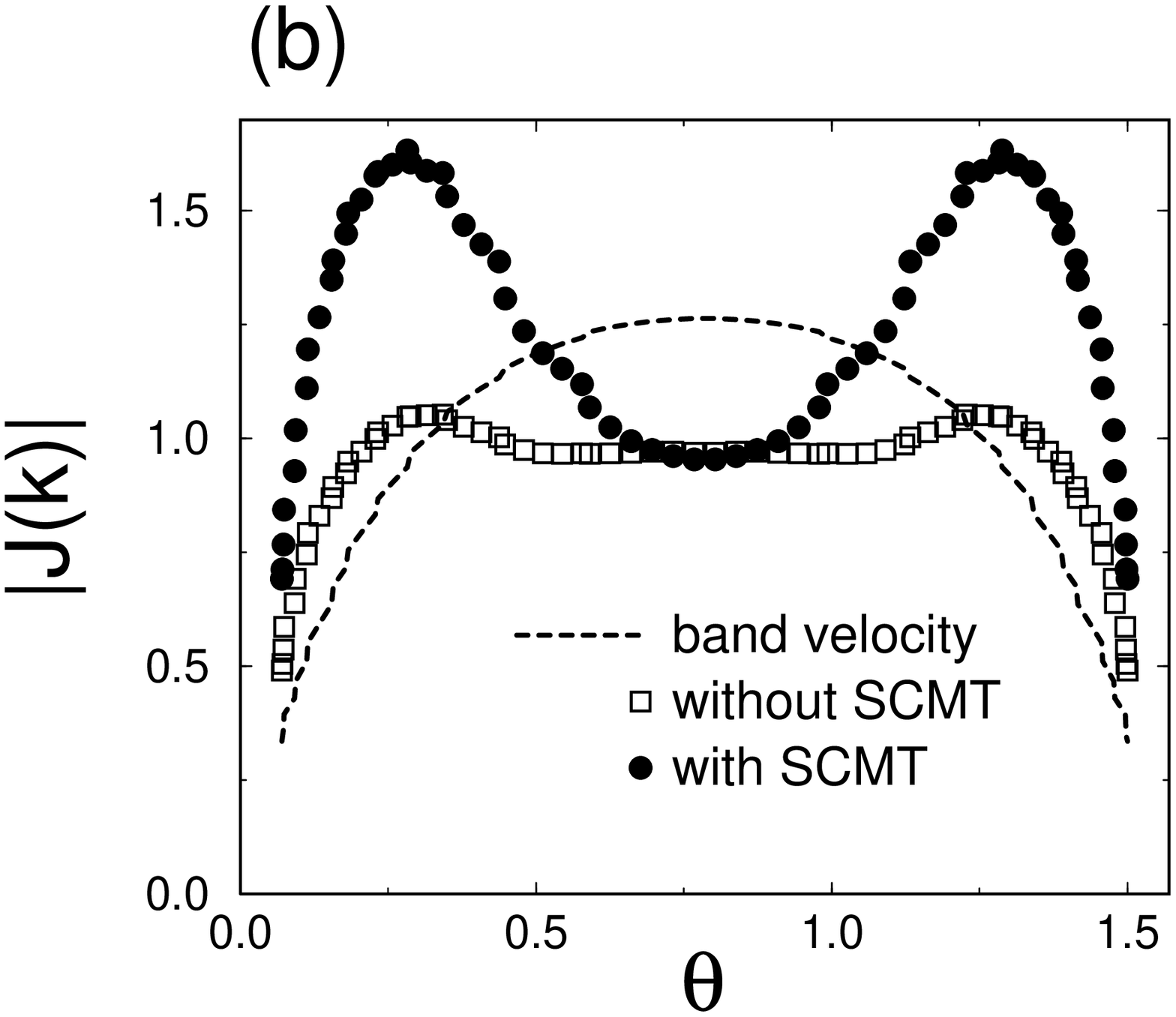}$$
    \caption{(a) The angle of the current vertex 
             $\varphi(\k)={\rm Arctan}(J_{x}(\k)/J_{y}(\k))$ 
             on the Fermi surface. (b) The absolute value $|J(\k)|$. 
             The squares and the circles correspond to the results 
             with and without the SCMT term, respectively.  
             The dashed line shows the results of the band velocity. 
             The parameters are chosen as $U=2.0$, $\delta=0.09$ and 
             $T=0.004$. 
             }    
  \end{center}
\end{figure}

 In addition to the above correction, we calculate 
the AL term (Fig. 13(b)).~\cite{rf:AL} 
 The AL term has divergent contribution 
in the vicinity of the critical point $T=T_{{\rm c}}$ and is interpreted 
as the conductivity carried by the fluctuating Cooper pairs.~\cite{rf:tinkham} 
 Therefore, this contribution is written as the 
superconducting part $\sigma_{{\rm s}}$ in contrast to the normal part 
$\sigma_{{\rm n}}$. 
 Here, we define the normal part $\sigma_{{\rm n}}$ as 
 the lowest order contribution with respect to $1/\tau(\k)$. 
 The expression of the AL term is obtained as follows, 
\begin{eqnarray}
  \sigma_{{\rm s}} & = & \frac{1}{\omega} 
                         {\rm Im} K_{{\rm AL}}(\omega)|_{\omega \rightarrow 0},
\\
  K_{{\rm AL}}(\omega_{n}) & = & (2 e)^{2} T \sum_{\q,\Omega_{n}} 
                           C_{x}(\q,{\rm i} \Omega_{n}:{\rm i} \omega_{n})^{2} 
\nonumber \\
& &\times
                           t(\q,{\rm i} \Omega_{n}) 
                           t(\q,{\rm i} \Omega_{n}+{\rm i} \omega_{n}), 
\\
  C_{x}(\q,{\rm i} \Omega_{n}:{\rm i} \omega_{n}) & = & 
                           T \sum_{\k,\omega_{m}} v_{x}(\k) 
                           \phi(\k,{\rm i} \omega_{m}) 
                           \phi^{*}(\k,{\rm i} \omega_{m}+{\rm i} \omega_{n}) 
\nonumber \\ 
     \times             G (\k,{\rm i} \omega_{m}) 
                           G (&\k&,{\rm i} \omega_{m}+{\rm i} \omega_{n})
                           G (\k-\q,{\rm i} \Omega_{n}-{\rm i} \omega_{m}). 
\end{eqnarray}
 The $\omega_{n}$-dependence of the three point vertex function 
$C_{x}(\q,{\rm i} \Omega_{n}:{\rm i} \omega_{n})$ is usually neglected 
because it gives only the less divergent contribution with respect to 
$1/t_{0}$. 
 Actually, this procedure corresponds to the exclusion of the normal part 
contribution of the AL term. 
 We confirmed numerically that this normal part contribution is small 
compared with that from the SCMT term. 
 Hereafter, we neglect the $\omega_{n}$-dependence in 
$C_{x}(\q,{\rm i}\Omega_{n}:{\rm i}\omega_{n})$ 
in order to maintain the consistency with the transverse conductivity. 
 This choice does not affect the following results, qualitatively. 
 The equations.(5.5-7) result in the following expression 
in the weak coupling limit, 
\begin{eqnarray}
 \sigma_{{\rm s}} =  \frac{e^{2}}{16} \frac{T}{T-T_{{\rm c}}}. 
\end{eqnarray}
 This is the well-known result in two dimension. 
 Here, we directly calculate the function $K_{{\rm AL}}(\omega_{n})$ 
on the imaginary axis, and obtain the conductivity 
$\sigma_{{\rm s}}$ by the analytic continuation. 
 The total conductivity $\sigma$ is obtained by the summation 
$\sigma = \sigma_{{\rm n}}+\sigma_{{\rm s}}$.

\begin{figure}[htbp]
  \begin{center}
   \epsfysize=6cm
    $$\epsffile{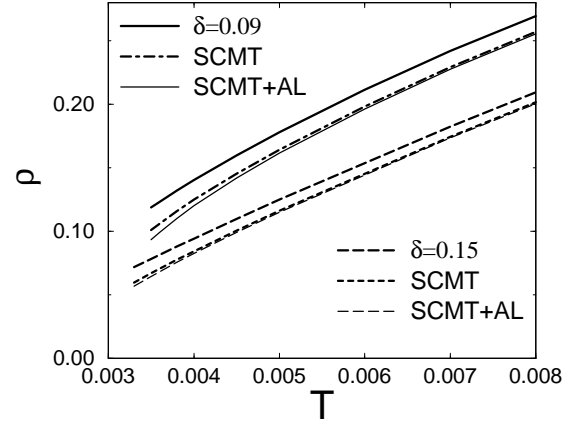}$$
    \caption{The effects of the vertex correction on the resistivity $\rho$. 
             The thick solid, dash-dotted and thin solid lines 
             are the results at $\delta=0.09$ with SPMT term, 
             with SPMT and SCMT terms and with SPMT, SCMT and AL terms, 
             respectively. 
             The thick long-dashed, dashed and thin long-dashed lines show the 
             same results at $\delta=0.15$. The interaction is fixed to 
             $U=2.0$. 
             }    
  \end{center}
\end{figure}

 We show the obtained results of the resistivity in Fig. 15. 
 Both the SCMT and AL terms decrease the resistivity, as is expected 
in the above discussions. 
 The effect of the SCMT term increases with decreasing the temperature. 
 This natural trend is indistinct in Fig. 15 because the absolute value of the
resistivity becomes small at the low temperature. 
 It is an important result that the SCMT term does not significantly affect 
the temperature dependence of the resistivity. 
 The effect of the SCMT term becomes more indistinct with hole-doping, 
which is a natural result because the SC fluctuations become weak.

 The other important result is that the AL term is almost negligible 
except for the narrow region near the critical point. 
 The contribution is sufficiently small even in the deeply critical region 
$T < T_{{\rm c}}^{\rm MF}$. 
 The divergent behavior of the AL term is not shown in our results because 
the calculation is carried out within the region $t_{0} > 0.01$.  
 (The numerical error becomes serious as the parameter $t_{0}$ becomes too 
small.)

 This is an expected behavior in case of the strong coupling 
superconductivity. 
 The AL term is expressed by the universal expression (eq. (5.8)) 
which does not depend on the parameters of the fluctuations, 
while the effect on the single particle properties is enhanced 
in the strong coupling superconductors. 
 In other words, the pseudogap occurs under the relatively large value of 
$t_{0}$ where the AL term is small. 
 Therefore, the pseudogap phenomena in the single particle properties 
dominate the contribution of the AL term at least in the strong coupling 
limit.

 The other important aspect is the ratio of the superconducting part 
$\sigma_{{\rm s}}$ to the normal part $\sigma_{{\rm n}}$. 
 The AL term becomes important when the absolute value of the normal part 
is small. 
 Since the normal part $\sigma_{{\rm n}}$ is large owing to the Cold spot, 
the superconducting part $\sigma_{{\rm s}}$ is relatively small in the 
High-$T_{{\rm c}}$ cuprates. 
 In other words, the AL term is a higher order term with respect to 
$1/\tau_{{\rm c}}$ which is a sufficiently small parameter 
(for example, $\leq 10^{-2}$ at $T=0.004$ and $\delta=0.09$). 
 Since the parameter $1/\tau_{{\rm c}}$ becomes even smaller 
with hole-doping, the AL term is more indistinct in the optimally-doped case 
(Fig. 15).

 In case of the strong coupling $s$-wave superconductors,  
the AL term should be much more important. This is because the pseudogap 
occurs on the whole Fermi surface and therefore the normal part 
$\sigma_{{\rm n}}$ is remarkably suppressed. 
 Thus, the momentum dependence of the $d$-wave superconductivity plays 
an important role in this stage.

 Here, we briefly comment on the ${\it c}$-axis transport. 
 The normal part of the conductivity $\sigma_{{\rm n}}$ is significantly small 
along the ${\it c}$-axis. However, the AL term is negligible even 
in this case. This is because the AL term is higher order with respect 
to the inter-layer hopping $t_{\perp}(\k)$, ${\it i.e.}$ 
$\sigma_{{\rm s}} \propto t_{\perp}^{4}$ while 
$\sigma_{{\rm n}} \propto t_{\perp}^{2}$.~\cite{rf:IAV,rf:dorin}  
 It is clear that the AL term is negligible when the two dimensionality 
is strong. 
 It is again noted that the qualitative difference between the 
${\it c}$-axis and in-plane transport is not attributed to this difference 
of the AL term,~\cite{rf:IAV,rf:dorin} but to the difference of 
the normal part (See the last of \S2.1).

\begin{figure}[htbp]
  \begin{center}
   \epsfysize=6cm
    $$\epsffile{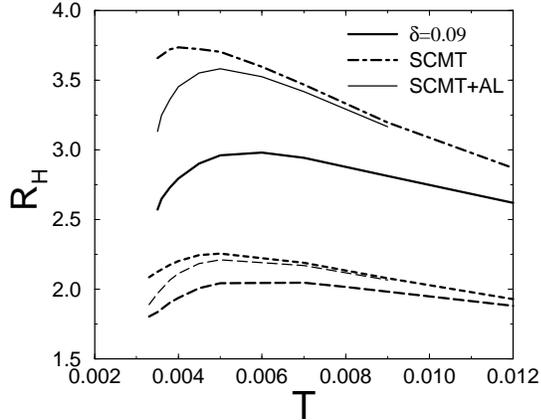}$$
    \caption{The effects of the vertex correction on the Hall coefficient 
             $R_{{\rm H}}$. The meanings of the lines are  
             the same as those in Fig. 15. 
             }    
  \end{center}
\end{figure}

 We show the results of the Hall coefficient in Fig. 16.  
 It is shown that the SCMT term enhances the Hall coefficient as is expected. 
 However, this effect is much smaller than the feedback effect discussed in 
\S4. 
 Therefore, the qualitatively the same behaviors are obtained,  
although the decrease of the Hall coefficient is reduced. 
 If we consider the AL term, the Hall coefficient decreases more clearly. 
 The effect of the AL term is more distinct in the Hall coefficient. 
 As a result, the qualitatively same results are obtained as those obtained 
in \S4. 
 In particular, the downward deviation of the Hall coefficient in the 
pseudogap state is robust. 
 The response of the Hall coefficient to the pseudogap is rather strong 
compared with that of the resistivity. 
 Thus, the discussion which is given before is confirmed 
by the numerical results.

 Here, we comment on the contribution of the AL term on the Hall 
conductivity.~\cite{rf:aronov,rf:ebisawa} 
 This contribution has been discussed in connection with the 
sign change of the Hall conductivity which is observed in the vicinity of 
the critical point.~\cite{rf:matsuda} 
 Generally speaking, the contribution is much smaller than that to the 
longitudinal conductivity $\sigma_{{\rm s}}$. The ratio is the order 
$T_{{\rm c}}/\varepsilon_{{\rm F}}$ and is estimated by the 
$\Omega$-dependence of the 
T-matrix.~\cite{rf:dorsey,rf:aronov,rf:ebisawa,rf:geshkenbein} 
 Although the ratio $T_{{\rm c}}/\varepsilon_{{\rm F}}$ is relatively  
large in High-$T_{{\rm c}}$ cuprates, it is still small of the 
order $\sim 10^{-1}$ in this calculation. 
 Therefore, the contribution to the Hall conductivity is 
expected to be negligible in the region of our interest.

\section{Conclusion and Discussions}

 In this paper, we have investigated the transport phenomena in the pseudogap 
state of High-$T_{{\rm c}}$ cuprates,  
 assuming that the pseudogap phenomena are caused by the strong 
SC fluctuations. 
 We previously developed a microscopic formalism which describes 
the SC fluctuations and the pseudogap phenomena 
by starting from the Hubbard model.~\cite{rf:yanaseFLEXPG} 
 We have applied the formalism to the calculation of the transport 
coefficients. 
 In this paper, the attention has been mainly focused on the in-plane electric 
transport, such as the resistivity $\rho$ and the Hall coefficient 
$R_{{\rm H}}$. A brief comment on the $c$-axis transport has been given.

 Since the conductivity is expressed by the two point correlation function,  
not only the single-particle Green function but also the four point vertex are 
necessary for the calculation. 
 First, we have performed the calculation in which the vertex correction 
arising from the spin fluctuations (SPMT term) is included. 
 We have shown in \S4 that the characteristic behaviors in the pseudogap 
state are well explained within this calculation. 
 The resistivity is slightly reduced by the SC fluctuations and  
deviates downward in the pseudogap 
state.~\cite{rf:ito,rf:takenaka,rf:odatransport,rf:mizuhashi}
 The Hall coefficient also deviates downward and shows a broad peak above 
$T_{{\rm c}}$.~\cite{rf:ito,rf:satoM,rf:haris,rf:mizuhashi,rf:xiong,rf:ong}
 We wish to point out that quite opposite behaviors are obtained 
if one considers only the SC fluctuations. 
 The correct results are obtained by considering the SC fluctuations and 
the AF spin fluctuations simultaneously. 
 Actually, the direct effects of the pseudogap and the feedback effects 
through the spin fluctuations compete with each other.
 Our calculation has quantitatively shown 
that the feedback effects exceed the direct effects, and 
the consistent results with experiments are obtained. 
 The above behaviors 
becomes clearer with decreasing hole-doping $\delta$ and/or increasing $U$. 
 These results are 
also consistent with the experimental results.

 It has been pointed out that the characteristic momentum dependence plays an 
essential role in the above argument. 
 One is the existence of the Hot spot and the Cold spot. 
 Another one is the $d$-wave form of the pseudogap. 
 Roughly speaking, the pseudogap phenomena occurs at the part of 
Fermi surface (Hot spot) which is not important for the electric transport 
from the beginning. 
 The easily flowing quasi-particles at the Cold spot are mainly affected 
by the spin fluctuations, rather than the SC fluctuations. 
 Therefore, the transport phenomena in the pseudogap state are dominated 
by the feedback effects. 
 If the superconductivity is the $s$-wave, or if the momentum dependent 
lifetime is neglected, 
the qualitatively inconsistent results are obtained.

 The above mentioned momentum dependence yields an outstanding character 
of the transport phenomena. While many of the interesting phenomena, 
such as the single particle's pseudogap, magnetic properties and so on, 
occur mainly at the Hot spot. On the other hand, the in-plane transport 
is mainly determined by the Cold spot. 
 This is why the in-plane transport phenomena show relatively weak change in 
 the pseudogap state, compared with the other properties. 
 The Hall coefficient has rather remarkable response to the pseudogap 
than the resistivity. 
 This difference has been explained by considering the SPMT term. 
 Thus, the characteristic responses to the pseudogap phenomena are 
systematically explained by our scenario.

 We have shown the results including the vertex correction arising from the 
SC fluctuations in \S5. 
 First, the lowest order term  (SCMT term) with respect to both $1/\tau(k)$ 
and the T-matrix has been estimated. 
 The SCMT term reduces the resistivity and enhances the Hall coefficient. 
However, since the vertex correction also include the $d$-wave form 
factor, the correction at the Cold spot is small while it is 
large at the Hot spot. 
 Therefore, the effects of the SCMT term are weak compared with the 
SPMT term, and the behaviors of the transport coefficients are not 
significantly affected. 

 In addition to the lowest order calculation, 
 we have estimated the AL term. 
 Our results show that the effect of the AL term appears only in the narrow 
region near the critical point. Therefore, this term is negligible in the 
wide temperature region in the pseudogap state. 
 This is simply because the parameter $1/\tau(k)$ is sufficiently small 
at the Cold spot and therefore the normal part contribution 
$\sigma_{{\rm n}}$ is large. 
 This is a characteristic result of the $d$-wave strong coupling 
superconductivity. 
 In the weak coupling theory of the SC fluctuations, 
the AL term or MT term is usually considered as the origin of the increasing 
conductivity. 
 However, the relative importance of the AL term is much reduced in the 
$d$-wave strong coupling superconductors, and the effects through the 
single particle and/or the magnetic properties become important. 
 Therefore, the downward deviation of the resistivity is not attributed to the 
AL term, but to the feedback effect or the unrelated effect.

 To summarize, the transport phenomena in the pseudogap state have been 
investigated on the basis of the microscopic calculation starting from 
the Hubbard model. The consistent understanding with the experimental results 
has been obtained. 
 Although there are still some issues to improve the approximation,  
for example, the calculation beyond the FLEX approximation,  
the essence of the electric transport in the pseudogap state has 
been explained in this paper.  
 It gives a strong evidence for the superconducting fluctuations as the origin 
of the pseudogap that the microscopic calculation gives a natural explanation 
of the transport phenomena which have essentially different characters 
from the other properties.

\section*{Acknowledgements}

 The author is grateful to Professor K. Yamada and Professor M. Ogata for 
fruitful discussions and . 
 Numerical computation in this work was partly carried out 
at the Yukawa Institute Computer Facility. 
 The present work was partly supported by a Grant-In-Aid for Scientific 
Research from the Ministry of Education, Science, Sports and Culture, Japan.


\begin{thebibliography}{9}
%
\bibitem{rf:yasuoka} 
H. Yasuoka, T. Imai and T. Shimizu: 
{\it Strong Correlation and Superconductivity} 
(Springer, Verlag, Berlin, 1989), p. 254. 
\bibitem{rf:NMR} For example,
H. Alloul, T. Ohno and P. Mendels: Phys. Rev. Lett. {\bf 63} (1989) 1700; 
W. W. Warren, R. E. Walstedt, G. F. Brennert, R. J. Cava, 
R. Tycko, R. F. Bell and G. Dabbagh: Phys. Rev. Lett. {\bf 62} (1989) 1193; 
M. Takigawa, A. P. Reyes, P. C. Hammel, J. D. Thompson, R. H. Heffner, 
Z. Fisk and K. C. Ott: Phys. Rev. B {\bf 43} (1991) 247; 
Y. Itoh, H. Yasuoka, Y. Fujiwara, Y. Ueda, T. Machi, I. Tomeno, K. Tai, 
N. Koshizuka and S. Tanaka: J. Phys. Soc. Jpn. {\bf 61} (1992) 1287;  
 M. H. Julien, P. Carretta, M. Horvati$\acute{{\rm c}}$: Phys. Rev. Lett. 
{\bf 76} (1996) 4238; 
Y. Itoh, T. Machi, S. Adachi, A. Fukuoka, K. Tanabe and 
H. Yasuoka: J. Phys. Soc. Jpn. {\bf 67} (1998) 312; 
K. Ishida, K. Yoshida, T. Mito, Y. Tokunaga, Y. Kitaoka, K. Asayama, 
Y. Nakayama, J. Shimoyama and K. Kishio: Phys. Rev. B {\bf  58} (1998) R5960. 
\bibitem{rf:neutron} 
J. Rossat-Mignod, L. P. Regnault, C. Vettier, P. Burlet, J. Y. Henry 
and G. Lapertot: Physica B {\bf 169} (1991) 58; 
J. Rossat-Mignod, L. P. Regnault, C. Vettier, 
P. Bourges, P. Burlet, J. Bossy, J. Y. Henry and G. Lapertot: 
Physica C {\bf 185\&189} (1991) 86. 
\bibitem{rf:iyereview}
The experimental results for the transport properties are reviewed in, 
Y. Iye: {\it Physical Properties of High Temperature Superconductors III}, 
ed. D. M. Ginsberg, (World Sci. Pub., Singapore, 1992) pp.285-361. 
\bibitem{rf:takagi} 
H. Takagi, T. Ido, S. Ishibashi, M. Uota, 
S. Uchida and Y. Tokura: Phys. Rev. B {\bf40} (1989) 2254; 
H. Takagi, B. Batlogg, H. L. Kao, J. Kwo, R. J. Cava, J. J. Krajewski and 
W. F. Peck, Jr.: Phys. Rev. Lett. {\bf 69} (1992) 2975. 
\bibitem{rf:kubo} 
Y. Kubo, Y. Shimakawa, T. Manako and H. Igarashi: 
Phys. Rev. B {\bf43} (1991) 7875. 
\bibitem{rf:ito} 
T. Ito, K. Takenaka and S. Uchida: Phys. Rev. Lett. {\bf 70} (1993) 3995.
\bibitem{rf:takenaka} 
K. Takenaka, K. Mizuhashi, H. Takagi and S. Uchida: 
Phys. Rev. B {\bf50} (1994)  6534. 
\bibitem{rf:mizuhashi} 
K. Mizuhashi, K. Takenaka, Y. Fukuzumi and S. Uchida: 
Phys. Rev. B {\bf 52} (1995) R3884. 
\bibitem{rf:odatransport}
M. Oda, K. Hoya, R. Kubota, C. Manabe, N. Momono, T. Nakano and M. Ido: 
Physica C {\bf 281} (1997) 135. 
\bibitem{rf:haris}
J. M. Haris, Y. F. Yan and N. P. Ong: Phys. Rev. B {\bf 46} (1992) 14293.
\bibitem{rf:satoM} 
T. Nishikawa, J. Takeda and M. Sato: J. Phys. Soc. Jpn. {\bf  62} (1993) 2568;
T. Nishikawa, J. Takeda and M. Sato: J. Phys. Soc. Jpn. {\bf  63} (1994) 1441;
J. Takeda, T. Nishikawa and M. Sato: Physica C {\bf 231} (1994) 293. 
\bibitem{rf:xiong} 
P. Xiong, G. Xiao and X. D. Wu: Phys. Rev. B {\bf 47} (1993) 5516. 
\bibitem{rf:hwang} 
H. Y. Hwang, B. Batlogg, H. Takagi, H. L. Kao, J. Kwo, R. J. Cava,
J. J. Krajewski and W. F. Peck, Jr: Phys. Rev. Lett. {\bf 72} (1994) 2636. 
\bibitem{rf:ong}
Z. A. Xu, Y. Zhang and N. P. Ong: cond-mat/9903123.
\bibitem{rf:homes} 
For example, 
C. C. Homes, T. Timusk, R. Liang, D. A. Bonn and W. H. Hardy: 
Phys. Rev. Lett. {\bf 71} (1993) 1645;
D. N. Basov, R. Liang, B. Dabrowski, D. A. Bonn, W. N. Hardy and T. Timusk: 
Phys. Rev. Lett. {\bf 77} (1996) 4090;
S. Tajima, J. Sch$\ddot{{\rm u}}$tzmann, S. Miyamoto, I. Terasaki, Y. Sato 
and R. Hauff: Phys. Rev. B {\bf 55} (1997) 6051. 
\bibitem{rf:momono} 
J. W. Loram, K. A. Mirza, J. M. Wade, J. R. Cooper and 
W. Y. Liang: Physica C {\bf 235\&240} (1994) 134; 
N. Momono, T. Matsuzaki, T. Nagata, M. Oda and M. Ido: 
to be published in J. Low. Temp. Phys. 
\bibitem{rf:renner} 
Ch. Renner, B. Revaz, J.-Y. Genoud, K. Kadowaki and 
\O. Fischer: Phys. Rev. Lett. {\bf 80} (1998) 149. 
\bibitem{rf:ARPES} 
H. Ding, T. Yokoya, J. C. Campuzano, T. Takahashi, 
M. Randeria, M. R. Norman,  T. Mochiku, K. Kadowaki and J. Giapintzakis:  
Nature. {\bf 382} (1996) 51; A. G. Loeser, Z. X. Shen, D. S. Dessau, 
D. S. Marshall, C. H. Park, P. Fournier and A. Kapitulnik: 
Science. {\bf 273} (1996) 325;
M. R. Norman, H. Ding, M. Randeria, J. C. Campuzano, 
T. Yokoya, T. Takeuchi, T. Takahashi, T. Mochiku, K. Kadowaki, P. Guptasarma 
and D. G. Hinks: Nature {\bf 392} (1998) 157.
\bibitem{rf:timusk}
T, Timusk and B. Statt: Rep. Prog. Phys. {\bf 62} (1999) 61. 
\bibitem{rf:randeriareview} 
C. A. R. S$\acute{{\rm a}}$ de Melo, M. Randeria and J. R. Engelbrecht: 
Phys. Rev. Lett. {\bf 71} (1993) 3202;  
M. Randeria: {\it Bose-Einstein Condensation}, ed. A. Griffin, D. Snoke and 
S. Stringari (Cambridge University Press, Cambridge, 1994);  
J. R. Engelbrecht, M. Randeria and C. A. R. S$\acute{{\rm a}}$ de Melo: 
Phys. Rev. B {\bf  55} (1997) 15153; 
M. Randeria: cond-mat/9710223 and references there in. 
\bibitem{rf:emery} 
V. J. Emery and S. A. Kivelson: Phys. Rev. Lett. {\bf 74} (1995) 3253.
\bibitem{rf:haussman} 
R. Haussmann: Phys. Rev. B {\bf 49} (1994) 12975. 
\bibitem{rf:stintzing} 
S. Stintzing and W. Zwerger: Phys. Rev. B {\bf 56} (1997) 9004. 
\bibitem{rf:micnas} 
R. Micnas, M. H. Pedersen, S. Schafroth, T. Schneider, 
J. J. Rodr$\acute{{\rm i}}$guez-N$\acute{{\rm u}}\tilde{{\rm n}}$ez and 
H. Beck: Phys. Rev. B {\bf 52} (1995) 16223. 
\bibitem{rf:geshkenbein} 
V. B. Geshkenbein, L. B. Ioffe and A. I. Larkin: 
Phys. Rev. B {\bf 55} (1997) 3173. 
\bibitem{rf:dahmpseudogap} 
T. Dahm, D. Manske and L. Tewordt: Phys. Rev. B {\bf55} (1997) 15274. 
\bibitem{rf:janko} 
B. Jank$\acute{{\rm o}}$, J. Maly and K. Levin: 
Phys. Rev. B {\bf 56} (1997) 11407; 
J. Maly, B. Jank$\acute{{\rm o}}$ and K. Levin: 
Physica C {\bf 321} (1999) 113. 
\bibitem{rf:koikegamiNSR} 
S. Koikegami and K. Yamada: J. Phys. Soc. Jpn. {\bf 67} (1998) 1114. 
\bibitem{rf:kobayashiNSR} 
A. Kobayashi, A. Tsuruta, T. Matsuura and Y. Kuroda: 
J. Phys. Soc. Jpn. {\bf  67} (1998) 2626. 
\bibitem{rf:dagotto} J. R. Engelbrecht, A. Nazarenko, M. Randeria and 
E. Dagotto: Phys. Rev. B {\bf 57} (1998) 13406. 
\bibitem{rf:yanasePG} 
Y. Yanase and K. Yamada: J. Phys. Soc. Jpn. {\bf  68} (1999) 2999. 
\bibitem{rf:jujo} 
T. Jujo and K. Yamada: J. Phys. Soc. Jpn. {\bf  68} (1999) 2198. 
\bibitem{rf:yanaseMG} 
Y. Yanase and K. Yamada: J. Phys. Soc. Jpn. {\bf  69} (2000) 2209.
\bibitem{rf:yanaseSC} 
T. Jujo, Y. Yanase and K. Yamada: J. Phys. Soc. Jpn. {\bf  69} (2000) 2240; 
Y. Yanase, T. Jujo and K. Yamada: J. Phys. Soc. Jpn. {\bf  69} (2000) 3664. 
\bibitem{rf:yanaseFLEXPG} 
Y. Yanase and K. Yamada: J. Phys. Soc. Jpn. {\bf  70} (2001) 1659. 
\bibitem{rf:kobayasi} 
A. Kobayashi, A. Tsuruta, T. Matsuura and Y. Kuroda: 
J. Phys. Soc. Jpn. {\bf  68} (1999) 2506; 
{\it ibid} {\bf  70} (2001) 1214. 
\bibitem{rf:onoda} S. Onoda and M. Imada: J. Phys. Soc. Jpn. {\bf  68} 
(1999) 2762; {\it ibid.} {\bf  69} (2000) 312; {\it Proc. Int. Workshop, 
Magnetic Excitations in Strongly Correlated Electrons}, 
J. Phys. Soc. Jpn. (2000) Suppl. B, p. 32. 
\bibitem{rf:koikegami} 
S. Koikegami and K. Yamada: J. Phys. Soc. Jpn. {\bf  69} (2000) 768;  
{\it ibid} {\bf  69} (2000) 1950. 
\bibitem{rf:metzner}
D. Rohe and W. Metzner: Phys. Rev. B {\bf 63} (2001) 224509. 
\bibitem{rf:AL} L. G. Aslamazov and A. I. Larkin: Fiz. Tverd. Tela. {\bf  10} 
(1968) 1104. [Sov. Phys. Solid State {\bf  10} (1968) 875.] 
\bibitem{rf:MT} K. Maki: Prog. Theor. Phys {\bf  40} (1968) 193;
R. S. Thompson:  Phys. Rev. B {\bf  1} (1970) 327. 
\bibitem{rf:Nozieres} 
P. Nozi$\grave{{\rm e}}$res and S. Schmitt-Rink: 
J. Low Temp. Phys. {\bf  59} (1985) 195; 
A. J. Leggett: {\it Modern Trends in the Theory of Condensed Matter}, 
ed. A. Pekalski and R. Przystawa (Springer-Verlag, Berlin, 1980). 
\bibitem{rf:tokumitu} 
S. Schmitt-Rink, C. M. Varma and A. E. Ruckenstein: 
Phys. Rev. Lett. {\bf63} (1989) 445; 
A. Tokumitu, K. Miyake and K. Yamada: 
Prog. Theor. Phys. Suppl. {\bf  106} (1991) 63. 
\bibitem{rf:hlubina} 
R. Hlubina and T. M. Rice: Phys. Rev. B {\bf51} (1995) 9253. 
\bibitem{rf:stojkovic} 
B. P. Stojkovi$\acute{{\rm c}}$ and D. Pines: 
Phys. Rev. Lett. {\bf 76} (1996) 811; 
B. P. Stojkovi$\acute{{\rm c}}$ and D. Pines: 
Phys. Rev. B {\bf 55} (1997) 8576. 
\bibitem{rf:yanaseTR} 
Y. Yanase and K. Yamada: J. Phys. Soc. Jpn. {\bf  68} (1999) 548. 
\bibitem{rf:kontani} 
H. Kontani, K. Kanki and K. Ueda: Phys. Rev. B {\bf  59} (1999) 14723. 
\bibitem{rf:kanki} 
K. Kanki and H. Kontani: J. Phys. Soc. Jpn. {\bf  68} (1999) 1614.
%
%
%
\bibitem{rf:bieri} 
J. B. Bieri, K. Maki and R. S. Thompson: Phys. Rev. B {\bf 44} (1991) 4709. 
\bibitem{rf:IAV} 
L. B. Ioffe, A. I. Larkin, A. A. Varlamov and L. Yu: 
Phys. Rev. B {\bf  47} (1993) 8936. 
\bibitem{rf:dorin} 
V. V. Dorin, R. A. Klemm, A. A. Varlamov, A. I. Buzdin and D. V. Livanov: 
Phys. Rev. B {\bf  48} (1993) 12951. 
\bibitem{rf:varlamov} 
A. A. Varlamov, G. Balestrino, E. Milani and D. V. Livanov: 
Adv. Phys. {\bf 48} (1999) 655. 
\bibitem{rf:ikeda}
R. Ikeda, T. Ohmi and T. Tsuneto: J. Phys. Soc. Jpn. {\bf  58} (1989) 1377;
{\it ibid}  {\bf  59} (1990) 1037; 
{\it ibid}  {\bf  60} (1991) 1051. 
\bibitem{rf:dorsey}
S. Ullah and A. T. Dorsey: Phys. Rev. B {\bf  44} (1991) 262. 
\bibitem{rf:aronov} 
A. G. Aronov and A. B. Rapoport: Mod. Phys. Lett. B {\bf  6} (1992) 1083; 
A. G. Aronov, S. Hikami and A. I. Larkin; Phys. Rev. B {\bf51} (1995) 3880.
\bibitem{rf:ioffe} 
L. B. Ioffe and A. J. Millis: Phys. Rev. B {\bf 58} (1998) 11631. 
%
%
%
\bibitem{rf:FLEX} 
N. E. Bickers, D. J. Scalapino and S. R. White: 
Phys. Rev. Lett. {\bf 62} (1989) 961; N. E. Bickers and D. J. Scalapino: 
Ann. Phys. (N.Y.) {\bf 193} (1989) 206. 
\bibitem{rf:baym} 
G. Baym and L. P. Kadanoff: Phys. Rev. {\bf  124} (1961) 287. 
\bibitem{rf:monthouxFLEX} 
P. Monthoux and D. J. Scalapino: Phys. Rev. Lett. {\bf  72} (1994) 1874. 
\bibitem{rf:paoFLEX} 
C.-H. Pao and N. E. Bickers: Phys. Rev. Lett. {\bf  72} (1994) 1870; 
Phys. Rev. B {\bf  51} (1995) 16310. 
\bibitem{rf:dahmFLEX} 
T. Dahm and L. Tewordt: Phys. Rev. Lett. {\bf  74} (1995) 793; 
Phys. Rev. B {\bf  52} (1995) 1297. 
\bibitem{rf:langerFLEX} 
M. Langer, J. Schmalian, S. Grabowski and K. H. Bennemann: 
Phys. Rev. Lett. {\bf  75} (1995) 4508. 
\bibitem{rf:deiszFLEX} 
J. J. Deisz, D. W. Hess and J. W. Serene: Phys. Rev. Lett. {\bf  76} (1996) 
1312.
\bibitem{rf:koikegamiFLEX} 
S. Koikegami, S. Fujimoto and K. Yamada: J. Phys. Soc. Jpn. {\bf  66} (1997) 
1438. 
\bibitem{rf:takimotoFLEX} 
T. Takimoto and T. Moriya: J. Phys. Soc. Jpn. {\bf  66} (1997) 2459;
{\it ibid} {\bf  67} (1998) 3570. 
\bibitem{rf:moriyaAD} 
T. Moriya and K. Ueda: Adv. Phys. {\bf  49} (2000) 555. 
%
\bibitem{rf:moriya} 
T. Moriya, Y. Takahashi and K. Ueda: J. Phys. Soc. Jpn. {\bf  59} (1990) 2905; 
K. Ueda, T. Moriya and Y. Takahashi: J. Phys. Chem. Solids {\bf  53} (1992) 
1515. 
\bibitem{rf:monthoux} 
P. Monthoux, A. V. Balatsky and D. Pines: 
Phys. Rev. B {\bf  46} (1992) 14803. 
%
\bibitem{rf:eliashberg}
G. M. $\acute{{\rm E}}$liashberg: Sov. Phys.-JETP 
{\bf14} (1962) 886. [J. Exptl. Theoret. Phys. (U.S.S.R) {\bf41} (1961) 410.]
\bibitem{rf:yamada} 
K. Yamada and K. Yosida: Prog. Theor. Phys. {\bf76} (1986) 621.
\bibitem{rf:okabe} 
T. Okabe: J. Phys. Soc. Jpn. {\bf 67} (1998) 2792; {\it ibid} 4178. 
\bibitem{rf:maebashi} 
H. Maebashi and H. Fukuyama: J. Phys. Soc. Jpn. {\bf 66} (1997) 3577; 
{\it ibid} {\bf 67} (1998) 242.
\bibitem{rf:kohno}
H. Kohno and K. Yamada: Prog. Theor. Phys. {\bf80} (1988) 623. 
\bibitem{rf:okanderson} O. K. Anderson, A. I. Liechtenstein, O. Jepsen and 
F. Paulsen: J. Phys. Chem. Solids. {\bf56} (1995) 1573. 
\bibitem{rf:rosch}
A. Rosch: Phys. Rev. Lett. {\bf82} (1999) 4280. 
\bibitem{rf:mesot} J. Mesot, M. R. Norman, H. Ding, M. Randeria, 
J. C. Campuzano, A. Paramekanti, H. M. Fretwell, A. Kaminski, T. Takeuchi, 
T. Yokoya, T. Sato, T. Takahashi, T. Mochiku and K. Kadowaki: 
Phys. Rev. Lett. {\bf83} (1999) 840. 
\bibitem{rf:tinkham} 
M. Tinkham:{\it Introduction to Superconductivity} (McGraw-Hill, 1975) Chap. 7.
%
%
%
\bibitem{rf:ebisawa} 
H. Ebisawa and H. Fukuyama: Prog. Theor. Phys. {\bf  46} (1971) 1042.
\bibitem{rf:matsuda} 
T. Nagaoka, Y. Matsuda, H. Obara, A. Sawa, T. Terashima, I. Chong, M. Takano 
and M. Suzuki: Phys. Rev. Lett. {\bf 80} (1998) 3594. 
%
\end{thebibliography}
\end{document}